\documentclass[12pt,preprint]{aastex}

\shorttitle{The On/Off Nature of SPI}
\shortauthors{Shkolnik, Bohlender \& Walker}

\begin{document}


\title{The On/Off Nature of Star-Planet Interactions\altaffilmark{1}}


\author{Evgenya~Shkolnik}
\affil{NASA Astrobiology Institute, University of Hawaii at Manoa\\ 2680 Woodlawn Drive, Honolulu, HI 96822}
\email{shkolnik@ifa.hawaii.edu}

\author{David A. Bohlender}
\affil{Herzberg Institute for Astrophysics, National Research Council of Canada\\
Victoria BC, Canada V9E 2E7}
\email{david.bohlender@nrc-cnrc.gc.ca}

\author{Gordon A.H.~Walker}
\affil{1234 Hewlett Place, Victoria, BC, V8S 4P7}
\email{gordonwa@uvic.ca}

\and

\author{Andrew Collier Cameron}
\affil{School of Physics and Astronomy, University of St Andrews, North Haugh, St Andrews, Fife KY16 9SS}
\email{acc4@st-andrews.ac.uk}

\altaffiltext{1}{Based on observations collected at the Canada-France-Hawaii Telescope operated by the National Research Council of Canada, the Centre National de la Recherche Scientifique of France, and the University of Hawaii.}

\begin{abstract}

Evidence suggesting an observable magnetic interaction between a star and its hot Jupiter ($P_{orb}$ $<$ 7 days, $a$ $<$ 0.1 AU,
$M_p$sin$i$ $>$ 0.2 M$_J$) appears as a cyclic variation of stellar activity synchronized to the planet's orbit. In this study,
we monitored the chromospheric activity of 7 stars with hot Jupiters using new high-resolution \'echelle spectra collected with
ESPaDOnS over a few nights in 2005 and 2006 from the CFHT. We searched for variability in several stellar activity indicators
(Ca II H $\lambda$3968, K $\lambda$3933, the Ca II infrared triplet (IRT) $\lambda$8662 line, H$\alpha$ $\lambda$6563 and He I
$\lambda$5876). HD~179949 has been observed almost every year since 2001. Synchronicity of the Ca II H \& K emission with the
orbit is clearly seen in four out of six epochs, while rotational modulation with $P_{rot}$=7 days is apparent in the other two
seasons. We observe a similar phenomenon on $\upsilon$ And, which displays rotational modulation ($P_{rot}$=12 days) in
September 2005, in 2002 and 2003 variations appear to correlate with the planet's orbital period. This on/off nature of
star-planet interaction (SPI) in the two systems is likely a function of the changing stellar magnetic field structure
throughout its activity cycle. Variability in the transiting system HD 189733 is likely associated with an active region
rotating with the star, however, the flaring in excess of the rotational modulation may be associated with its hot Jupiter. As
for HD 179949, the peak variability as measured by the mean absolute deviation (MAD) for both HD 189733 and $\tau$ Boo leads the
sub-planetary longitude by $\sim70^{\circ}$. The tentative correlation between this activity and the ratio of $M_{p}$sin$i$ to
the planet's rotation period, a quantity proportional to the hot Jupiter's magnetic moment, first presented in Shkolnik et
al.~2005 remains viable. This work furthers the characterization of SPI, improving its potential as a probe of extrasolar
planetary magnetic fields.

\end{abstract}


\section{Introduction}\label{intro}

Observations and theory demonstrate that star-planet interaction (SPI) is a complex, yet potentially very informative probe of extrasolar planetary magnetic fields. In Shkolnik et al.~(2003, 2005a), we reported on planet-induced chromospheric activity on two stars, HD 179949 and $\upsilon$~And apparent from the night-to-night modulation of the Ca II H \& K chromospheric emission phased with the hot Jupiter's orbit. The modulation was indicative of a magnetic rather than tidal interaction (Cuntz et al.~2000), such that the period of the observed stellar activity correlated with the planet's orbital period $P_{orb}$, rather than $P_{orb}/2$. Ample observational evidence of tidal and magnetic interactions exists in the exaggerated case of the RS Canum Venaticorum (RS CVn) stars, which are tightly-orbiting binary systems consisting of two chromospherically active late-type stars (e.g.~Glebocki et al.~1986, Catalano et al.~1996, Shkolnik et al.~2005b ).

Although efforts to observe variable radio emission from the stars with hot Jupiters have not yet been successful (e.g.~Lazio \& Farrell 2007, George \& Stevens 2007), there have been several additional observations that support the existence of SPI. Photometric observations by the MOST space telescope of several hot-Jupiter systems, including HD 179949 and $\tau$~Boo, suggest that stellar surface activity in the form of active spots may be induced by the giant planet (Walker et al.~2006, 2007). 
Also, Saar et al.~(2006) recently reported a possible detection of planet-induced X-ray emission from the HD 179949 system corresponding to $\sim$30\% increase in X-ray flux over quiescent levels coincident with the phase of the Ca II enhancements at $\phi_{orb}$$\sim$0.8.  A statistical analysis by Kashyap et al.~(2006) suggests that the X-ray flux from stars with hot Jupiters is on average $\gtrsim$ 3 times greater than stars with planets at larger orbital distances, presenting further evidence that close-in giant planets have a measurable effect on the activity of the parent star.  

One scenario of magnetospheric interaction proposes that the planet induces reconnection events as it travels through the large stellar magnetic loops (Cuntz et al.~2000, Ip et al.~2004), implying that the resulting activity should depend on the star's magnetic field, the planet's magnetic field and the orbital distance with respect to the Alfv\'en radius of the host star ($\sim$10 stellar radii). Novel research of the magnetic field topology of hot Jupiter host stars is underway (Catala et al.~2007, Moutou et al.~2007) using Zeeman Doppler Imaging (ZDI), which hopes to contribute to a more detailed understanding of SPI. 

A detection of a magnetic field of a hot Jupiter would 1) provide a constraint on the rapid hydrodynamic escape of its atmosphere (Vidal-Madjar et al.~2003, 2004) which could affect the planet's structure and evolution, 2) present implications for the planet's internal structure, and 3) shed light on the mass-radius relationship of the known transiting planets (Pont et al.~2005, Bakos et al.~2006). 
Although the internal magnetic fields of hot Jupiters are expected to be weaker than Jupiter's due to probable tidal locking and slower spin rates (Sanchez-Lavega 2004, Griessmeier et al.~2004), Olson \& Christensen (2006) calculated that the magnetic field of a planet with even a tenth of Jupiter's rotation rate would still have a strong dipole moment, when reasonably assuming that the convection is not highly modified by the rotation rate.
Also, the fact that both hot and very hot Jupiters, such as HD 209458~b and OGLE-TR-56~b, are detected at all means that they must have strong enough magnetic fields to balance the extreme stellar irradiation and CME plasma pressure to prevent destructive atmospheric erosion (Khodachenko et al.~2007). 

We seek to probe hot Jupiter magnetic fields in order to understand their formation and evolution. SPI potentially offers an indirect way to detect, and with future modeling and observations, measure planetary magnetic fields.

It is reasonable to assume that any magnetic interaction would be greatest in the outermost layers of the star, namely the chromosphere, transition region and the corona due to their proximity to the planet, low density, and non-radiative heat sources. 
With the commissioning of CFHT's high-resolution \'echelle spectrograph, ESPaDoNS, we are able to include several stellar activity indicators in our analysis to observe the interaction as a function of atmospheric height in order to model the energy transfer and dissipation mechanisms of this phenomenon. ESPaDOnS' wavelength coverage allows simultaneous monitoring of the Ca II infrared triplet (IRT, lower chromosphere), H$\alpha$, Ca II H, K (middle chromosphere), and He I D$_3$ (upper chromosphere). 
 
Our program stars have planets with orbital periods between 2.2 and 7.1 days, eccentricities $\approx$~0 and semi-major axes $< 0.08$ AU.  These systems offer the best chance of observing upper atmospheric heating.  Of the seven systems we observed with ESPaDoNS, $\tau$~Boo, HD~179949, $\upsilon$ And and HD~209458 have been observed previously in our CFHT/Gecko campaign.  The first results from 2001 and 2002 observations, including the first evidence of planet-induced magnetic heating of HD~179949, were published in Shkolnik et al.~(2003, 2005a).  We later extended the experiment at the Very Large Telescope (VLT) to include five southern targets. The three new systems monitored in this study are HD 217107, HD 149143 and HD 189733. The system parameters for the ESPaDOnS program stars are listed in Table 1 along with our standard 61~Vir.

In this paper, we present new \'echelle spectra and compare with those of previous years, bringing to light a broader understanding of stellar activity, its cycles, and SPI. 
The details of our observations and data reduction are outlined in Section~\ref{spectra}. In Section~\ref{caII} we discuss our analysis and results of the Ca II K measurements including long-term, short-term and rotational modulation. Comparisons with other activity indicators are made in Section~\ref{other}. 

\clearpage
\begin{deluxetable}{ccccccccccccl}\label{targets}

\tabletypesize{\footnotesize}
\tablecaption{Stellar and Orbital Parameters \label{starpars_01}}
\tablewidth{0pt}
\tablehead{
\colhead{~ Star} &
\colhead{SpT} &
\colhead{$v$sin$i$} &
\colhead{$P_{rot}$} &
\colhead{$P_{orb}$\tablenotemark{a}} &
\colhead{$M_{p}$sin$i$\tablenotemark{a}} &
\colhead{$a$\tablenotemark{a}} &
\colhead{$\langle$K$\rangle$\tablenotemark{b}} &
\colhead{$\langle$K$\arcmin\rangle$\tablenotemark{c}} &
\colhead{$\langle$MADK$\rangle$\tablenotemark{d}} &
\colhead{He I EW} &
\\
\colhead{} &
\colhead{} &
\colhead{(km s$^{-1}$)} &
\colhead{(days)} &
\colhead{(days)} &
\colhead{($M_{J}$)} &
\colhead{(AU)} &
\colhead{(\AA)} &
\colhead{(\AA)} &
\colhead{(\AA)} &
\colhead{(m\AA)} &
}

\startdata

$\tau$~Boo  &   F7~IV 	&  	14.8$\pm$0.3  &   3.2\tablenotemark{e}   &3.31  &4.4   & 0.046 &  0.336 & 0.184 &  0.0019 & 27\\
HD~179949   &   F8~V  	&  	6.3$\pm$0.9   &  7\tablenotemark{f}   &3.09  &0.98   & 0.045 &  0.369 & 0.186&  0.0022\tablenotemark{j} & 17\\
HD~209458   &   G0~V  	&  	4.2$\pm$0.5   &  16\tablenotemark{g}  & 3.53  &0.69\tablenotemark{h}   & 0.045 &  0.195 & 0.078	&  0.0009 & $\lesssim$3\\
$\upsilon$ And	&	F7 V	&	9.0$\pm$0.4 & 12\tablenotemark{e,k}	&	4.618	&	0.71	&	0.059	&	0.254	&	0.091	&	0.0016 & --\\
HD~189733  &	K1 V & 	2.92$\pm$0.22 & 11.7\tablenotemark{i}   &2.22  &1.15\tablenotemark{h}    &0.031   & 1.337 & 1.231 & 0.0044\tablenotemark{j} & 35\\
HD~217107 	& 	G8 IV  	&  	9.0$\pm$0.4   &  39\tablenotemark{k}    & 7.13  &1.35   & 0.075 &  0.160 & 0.075	&  0.0007 & $<$4\\
HD~149143    &  G0 IV	&  	3.9$\pm$1 	&??     &4.09   &1.36   &0.052\tablenotemark{r} & 	0.342  & 0.144 & 0.0014 & 10\\
61~Vir   	&  	G5~V  	&  	2.2\tablenotemark{m}      & 	33\tablenotemark{k}   &   ---  &   ---   &  --- &    0.182  & 0.083 & 0.0008 & $<$4\\

\enddata

\tablenotetext{a}{Published orbital solutions: $\tau$ Boo $-$ Butler et al.~1997, HD 179949 $-$ Tinney et al.~2001, HD 209458 $-$ Charbonneau et al.~1999, $\upsilon$ And $-$ Butler et al.~1997, HD 189733 $-$ Bouchy et al.~ 2005, HD 217107 $-$ Fischer et al.~1998, HD~149143 $-$ De Silva et al.~2006}
\tablenotetext{b}{Total integrated intensity of the mean normalized Ca II K core. These values are relative to the normalization points near 3930 and 3937 \AA\/ at 1/3 of the pseudo-continuum at 3950 \AA.}
\tablenotetext{c}{We subtracted the photospheric emission from $\langle$K$\rangle$ in order to measure the mean integrated chromospheric emission $\langle$K$\arcmin\rangle$ using data from Wright et al.~(2004). (See text for more details.)}
\tablenotetext{d}{Average integrated `intensity' of the mean absolute deviation (MAD) of the K residuals, per observing run}
\tablenotetext{e}{Henry et al.~2000}
\tablenotetext{f}{This work and Wolf \& Harmanec 2004}
\tablenotetext{g}{Mazeh et al.~2000}
\tablenotetext{h}{Transiting system}
\tablenotetext{i}{Croll et al.~2007}
\tablenotetext{j}{Value were corrected to remove geometric (rotational and/or planetary) modulation of an active region on the star. For HD 189733, the non-corrected value is 0.0098 and for HD 179949, 0.0063.}

\tablenotetext{k}{Wright et al.~2004}

\end{deluxetable}

\begin{deluxetable}{ccccccccccl}\label{mags}

\tabletypesize{\footnotesize}
\tablecaption{Observations\label{starpars_02}}
\tablewidth{0pt}
\tablehead{
\colhead{~ Star} &
\colhead{$U$} &
\colhead{$B$} &
\colhead{$V$} &
\colhead{Exposures\tablenotemark{a}} &
\colhead{$S/N$\tablenotemark{b}} &
\colhead{$S/N$\tablenotemark{b}} &
\\
\colhead{} &
\colhead{} &
\colhead{} &
\colhead{} &
\colhead{$t \times n \times N$} &
\colhead{at 3950 \AA} &
\colhead{at 8710 \AA} &
}

\startdata

$\tau$ Boo	&	5.02	&	4.98	&	4.50	&	120 $\times$ 10 $\times$ 9\tablenotemark{c}	&	630	&	1640	\\
HD 179949	&	6.83	&	6.76	&	6.25	&	660 $\times$ 5 $\times$ 7\tablenotemark{e}	&	440	&	1270	\\
HD 209458	&	8.38	&	8.18	&	7.65	&	1800 $\times$ 4 $\times$ 4	&	350	&	960	\\
$\upsilon$ And & 	4.69	& 	4.63	&		4.09	&	180 $\times$ 120 $\times$ 4\tablenotemark{e}	&	2800 &	7290\\
HD 189733	&	$-$	&	8.60	&	7.67	&	1800 $\times$ 4 $\times$ 4	&	280	&	1260	\\
HD 217107	&	7.33	&	6.90	&	6.18	&	600 $\times$ 5 $\times$ 3	&	350	&	1340	\\
HD 149143	&	$-$	&	8.53	&	7.90	&	1800 $\times$ 4 $\times$ 4	&	320	&	1000	\\
61 Vir	&	5.71	&	5.45	&	4.74	&	120 $\times$ 7 $\times$ 5	&	450	&	1480	\\

\enddata

\tablenotetext{a}{$t$ = Exposure time in seconds, $n$ = number of exposures per night, $N$ = number of nights during the June 2006 ESPaDOnS observing run}
\tablenotetext{b}{Typical nightly S/N per 0.022-\AA\/ pixel}
\tablenotetext{c}{Spectra from three of the nine nights were observed in ESPaDOnS' `spectropolarimetry' mode.}
\tablenotetext{d}{Four nights in June 2006 and three nights in September 2005}  
\tablenotetext{e}{All these data were acquired in the `spectropolarimetry' mode in September 2005. The extremely high S/N was a requirement of the partner program to search for linear polarization. (Collier Cameron et al., in prep.)}

\end{deluxetable}
\clearpage

\section{The spectra}\label{spectra}

The observations were made with the 3.6-m Canada-France-Hawaii Telescope (CFHT) on 7 nights in
September 2005 and 9 nights in 2006 June. We used ESPaDOnS (\'Echelle SpectroPolarimetric Device for
the Observation of Stars), which is fiber fed from the Cassegrain to coud\'e
focus where the fiber image is projected onto a Bowen-Walraven slicer at the spectrograph entrance. With a 79 gr/mm grating
and a 2048$\times$4608-pixel CCD detector, ESPaDOnS' `star-only' mode records the full spectrum
over 40 grating orders covering 3700 to 10400 \AA\/ at a spectral resolution R of $\approx$80,000. The four nights of observations of $\upsilon$ And (18, 19, 23, 24 September 2005) and three nights of the $\tau$ Boo data (16, 17, 18 June 2006) were taken in ESPaDOnS' `spectropolarimetry' mode at R of 68,000.

The data were reduced using {\it Libre Esprit}, a fully automated reduction package provided for the
instrument and described in detail by Donati et al.~(1997, 2007\footnote{Also see
http://www.cfht.hawaii.edu/Instruments/Spectroscopy/Espadons/Espadons\_esprit.html.}). Each stellar
exposure is bias-subtracted and flat-fielded for pixel-to-pixel sensitivity variations. After
optimal extraction, the 1-D spectra are wavelength calibrated with several Th/Ar arcs taken
throughout the night. Finally the spectra are divided by a flat-field response and then the
continuum is normalized. Heliocentric velocity corrections are applied as well as small velocity
corrections ($<$ 100 m~s$^{-1}$) to account for instrumental effects using the telluric lines.

The final spectra were of high S/N reaching $\approx$ 130 per pixel (880 \AA$^{-1}$) in the H \& K
emission core, 400 per pixel (2700 \AA$^{-1}$) in the pseudo-continuum near 3950 \AA, and about 3
times higher near the Ca II IRT. Spectra with comparable S/N were taken of 61 Vir, a G5~V star
known not to have close-in giant planets, plus the hot standard HR 5511 (SpT = A0V) for telluric
line correction. Table 2 lists the program stars, including their magnitudes, exposure times, and
typical S/N.

\begin{figure}
\epsscale{.80}
\plotone{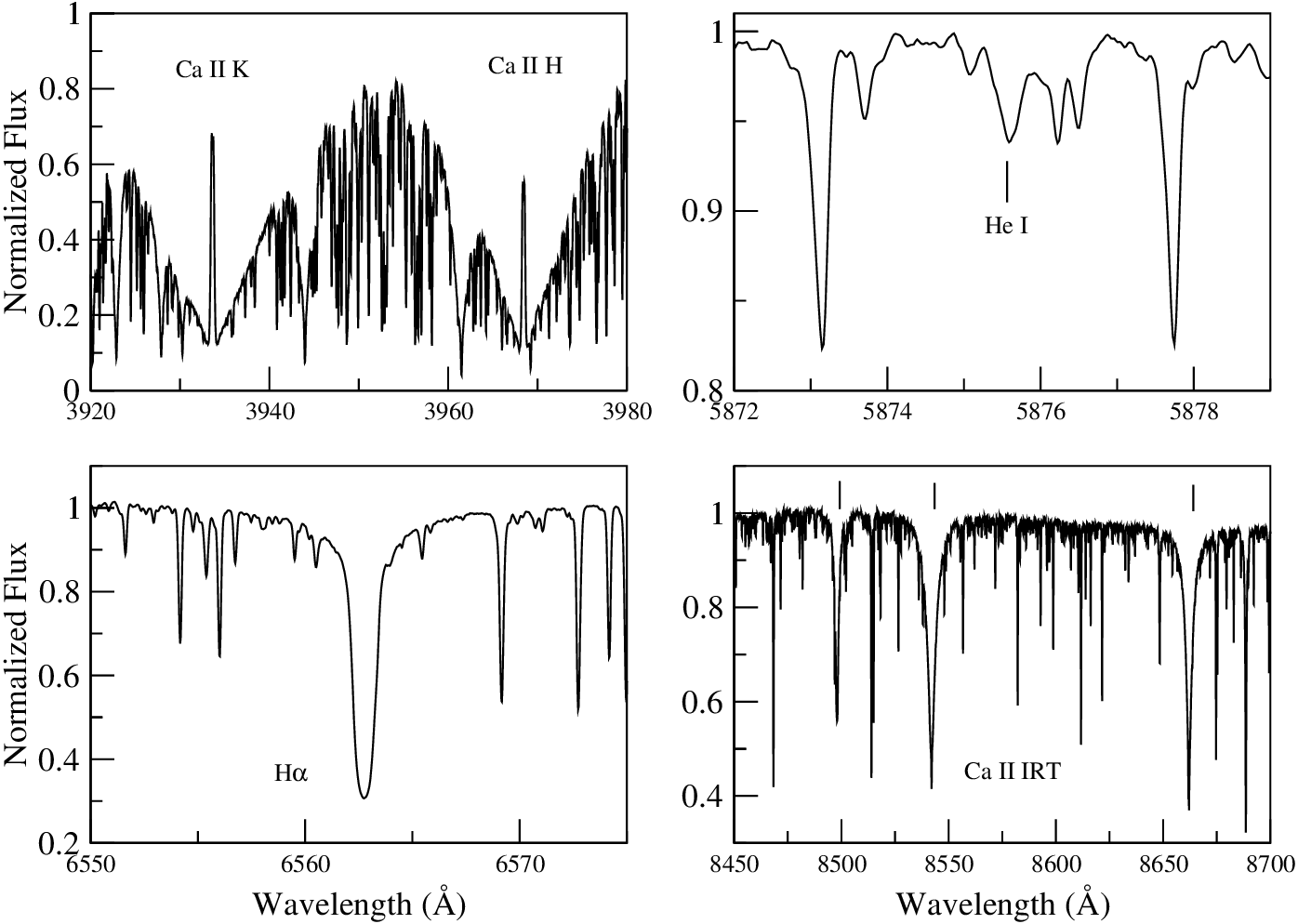}
\caption{Selected regions of a normalized spectrum of HD 189733 identifying key stellar activity diagnostics. 
\label{hd189733_spec}}
\end{figure}

All further processing and analysis were performed with standard IRAF (Image Reduction and Analysis
Facility) routines.\footnote[1]{IRAF is distributed by the National Optical Astronomy Observatories,
which is operated by the Association of Universities for Research in Astronomy, Inc.~(AURA) under
cooperative agreement with the National Science Foundation.} Differential radial velocity
corrections were applied to each stellar spectrum using IRAF's {\it fxcor} and {\it rvcorrect}
routines. Representative spectra near the key stellar activity indicators are shown in
Figure~\ref{hd189733_spec} for HD~189733.

\section{Measuring chromospheric activity}\label{caII}

\begin{figure}
\epsscale{.70}
\plotone{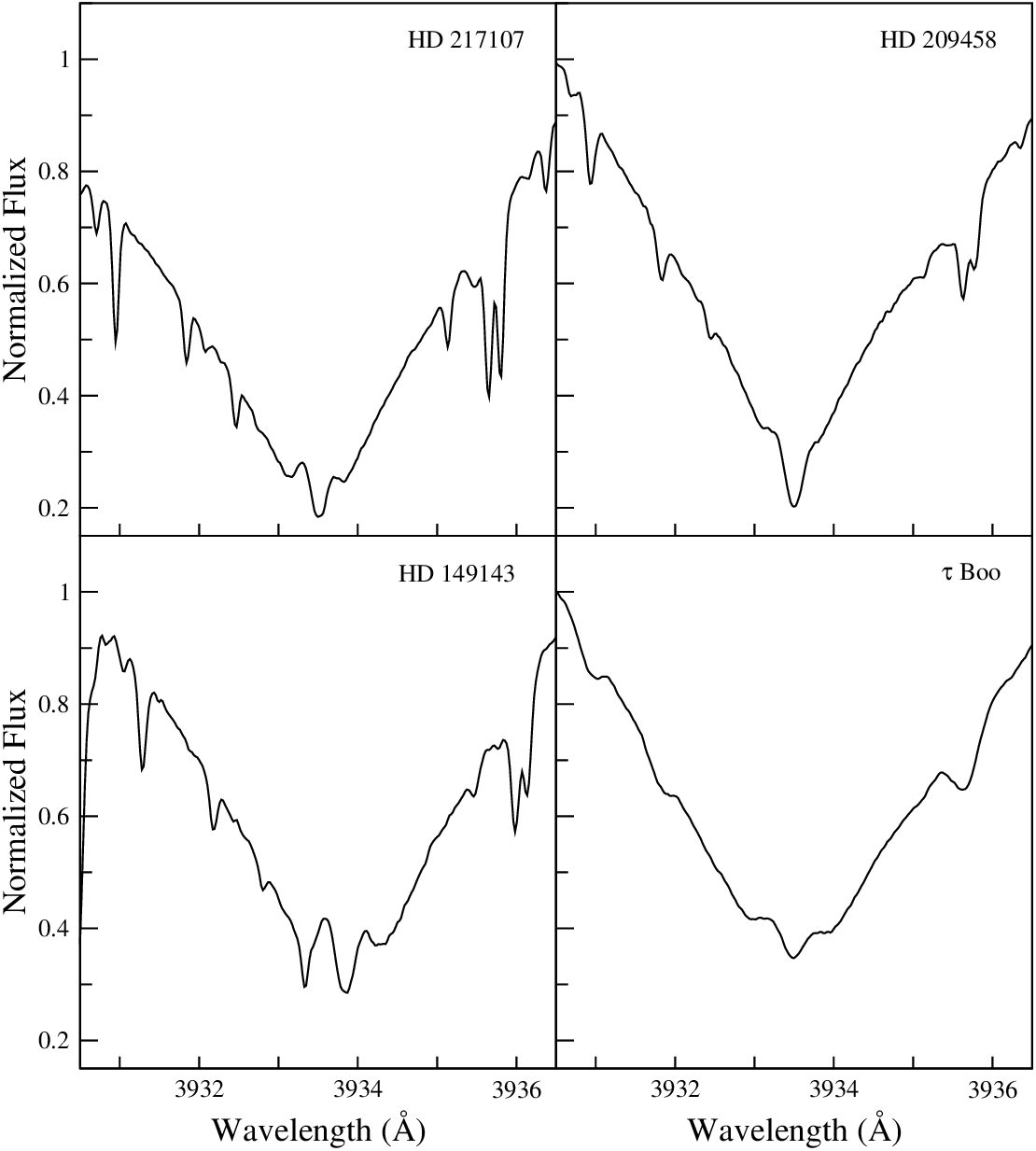}
\caption{The mean normalized Ca II K cores for four of the six program stars. That of HD 179949, HD 189733 and $\upsilon$ And are shown in Figures~\ref{hd179949_diffs}, \ref{hd189733_diffs} and \ref{upsAnd_diffs}, respectively.
\label{Kcores}}
\end{figure}

\begin{figure}
\epsscale{.70}
\plotone{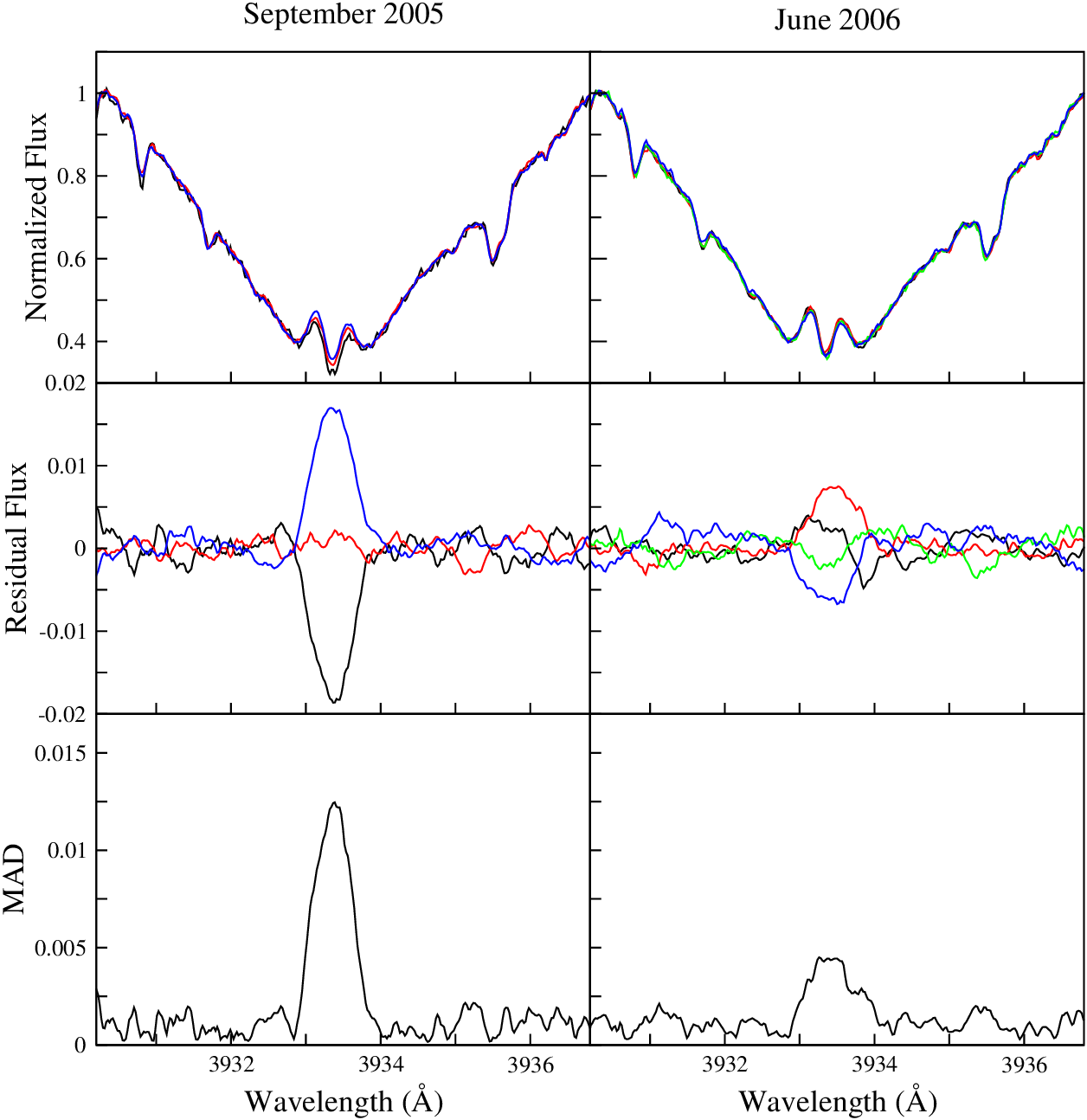}
\caption{Top: The mean normalized Ca II K emission of HD 179949 on the three nights observed in September 2005 and the four nights in June 2006.  Middle: The residuals relative to their respective means. Bottom: The mean absolute deviation (MAD) of the residuals. 
\label{hd179949_diffs}}
\end{figure}

\begin{figure}
\epsscale{0.5}
\plotone{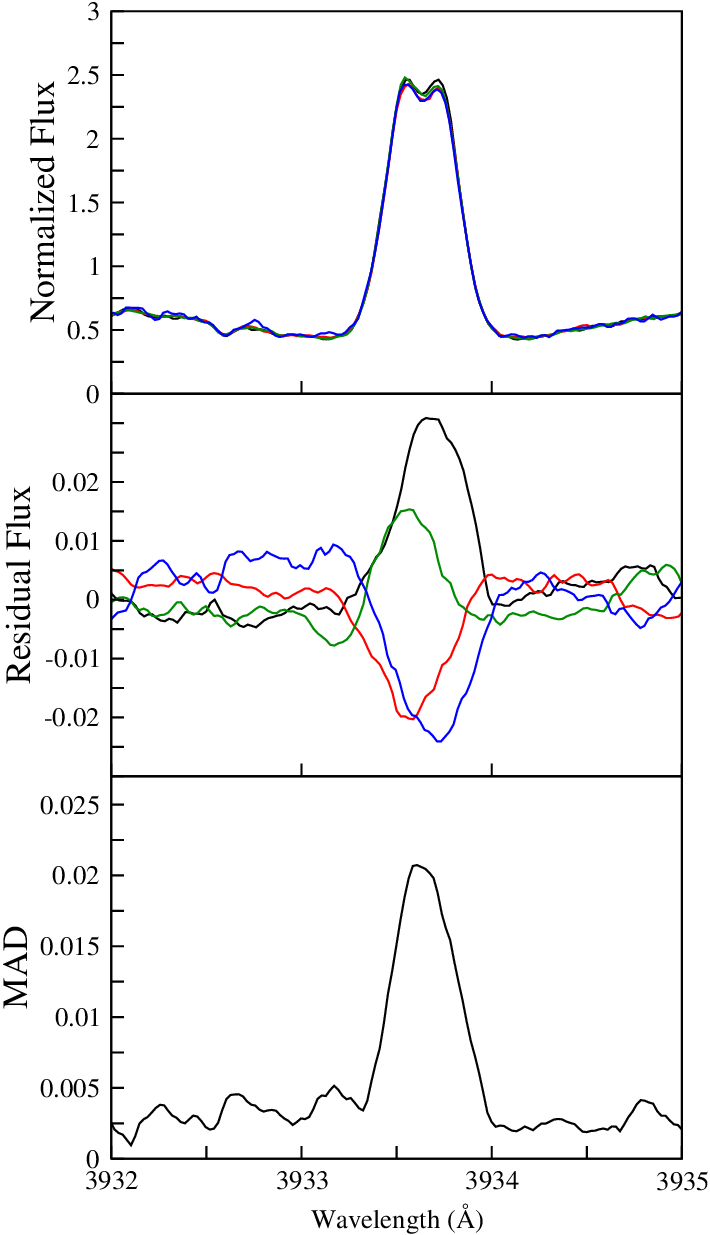}
\caption{As in Figure~\ref{hd179949_diffs} for HD 189733 observed in June 2006.
\label{hd189733_diffs}}
\end{figure}

\begin{figure}
\epsscale{0.5}
\plotone{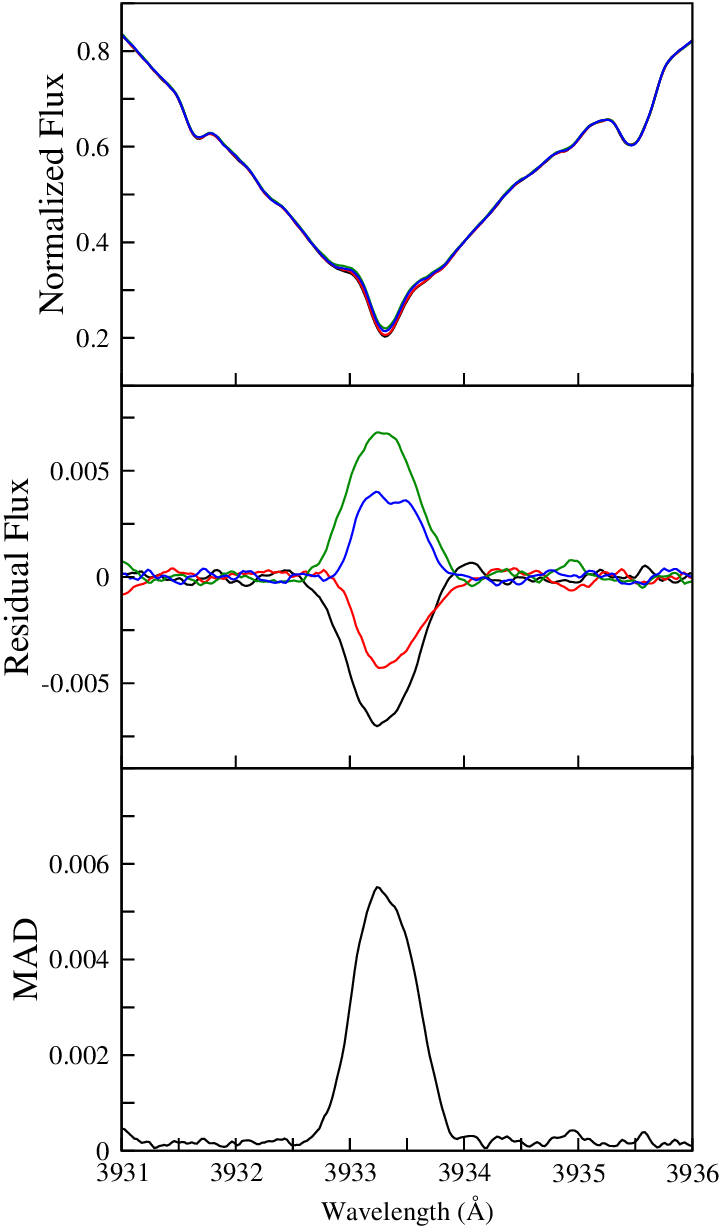}
\caption{As in Figure~\ref{hd179949_diffs} for $\upsilon$ And observed in September 2005. 
\label{upsAnd_diffs}}
\end{figure}

The very strong Ca II H and K photospheric absorption lines suppress the local stellar continuum making it difficult to normalize each spectrum
consistently. The normalization level was set at 0.3 of the flux at 3950~\AA\/ centered on the H and K lines. Therefore the wavelengths were
constant for all spectra of a given star, though they varied slightly from star to star due to variations in spectral type. The
$\approx$7-\AA\/ spectral range was chosen to isolate the H and K reversals.
This window is wide enough that a few photospheric absorption features appear to test for general
stability. To normalize each sub-spectrum, the end points were set to 1 and fitted with a straight line. The mean Ca II K cores for four of
the program stars are shown in Figure~\ref{Kcores} with those for HD~179949, HD~189733 and $\upsilon$ And in Figures~\ref{hd179949_diffs} $-$
\ref{upsAnd_diffs}.

The spectra were grouped by date and a nightly mean was computed for each of the lines. Other than
the active HD~189733 (see Section~\ref{hd189733}), all stars observed had non-varying K
emission at the $\lesssim$ 0.001 level on average on a given night. This is a result of S/N variations
and intranight (short time-scale) chromospheric activity.

We used nightly residuals from the average stellar spectrum to measure the chromospheric activity within the reversals. Each residual spectrum had a broad, low-order curvature removed. The residuals of the normalized spectra (smoothed by 17 pixels) were used
to compute the mean absolute deviation (MAD = $N^{-1}\Sigma|data_{i}-mean|$ for $N$
spectra), a measure of overall variability within the span of the observing run. The Ca II K MAD spectrum and the nightly residuals used to generate it for HD~179949, HD~189733 and $\upsilon$ And are displayed in Figures~\ref{hd179949_diffs} $-$ \ref{upsAnd_diffs}.

The analysis presented in this section consists of only Ca II K emission measurements for several reasons: 1) The broad, deep photospheric absorption of the Ca II K line allows the chromospheric emission to be seen at higher contrast as compared to H$\alpha$ and the Ca II IRT where chromospheric emission merely fills in the absorption core. 2) Extensive studies by the Mount Wilson group and Wright et al. (2004) allow us to isolate the average chromospheric emission $\langle$K$\arcmin$$\rangle$ by correcting for the photospheric contribution to our measurements ($\langle$K$\arcmin$$\rangle$ = $\langle$K$\rangle$$-$$\langle$K$_{phot}$$\rangle$, see Section 3.3.2 of Shkolnik et al.~2005a for more details.) This makes for a more accurate comparison of the stars in the sample whose spectral types vary from F7 to K1. 3) Previous CFHT/Gecko spectra consisted of only a single order containing Ca II H \& K and it is useful to make comparisons between the data sets. 4) Lastly, there are no telluric features or blended lines to contaminate the spectra, as is the case for the other indicators, which are discussed in Section~\ref{other}. 

\subsection{HD 179949 \& $\upsilon$ And: Evidence of the on/off nature of SPI}\label{induced}

When monitoring chromospheric emission, stellar activity may be modulated by the star's rotation, planetary motion in the case of SPI, or a combination of both. The orbital periods of the planets are well known and uniquely established by the PRV and transit discovery methods, but the rotation periods of the stars are much harder to determine in part due to stellar differential rotation. For studies of SPI, differentiating between rotational and orbital modulation of the chromospheric emission is key.

\begin{figure}
\epsscale{.80}
\plotone{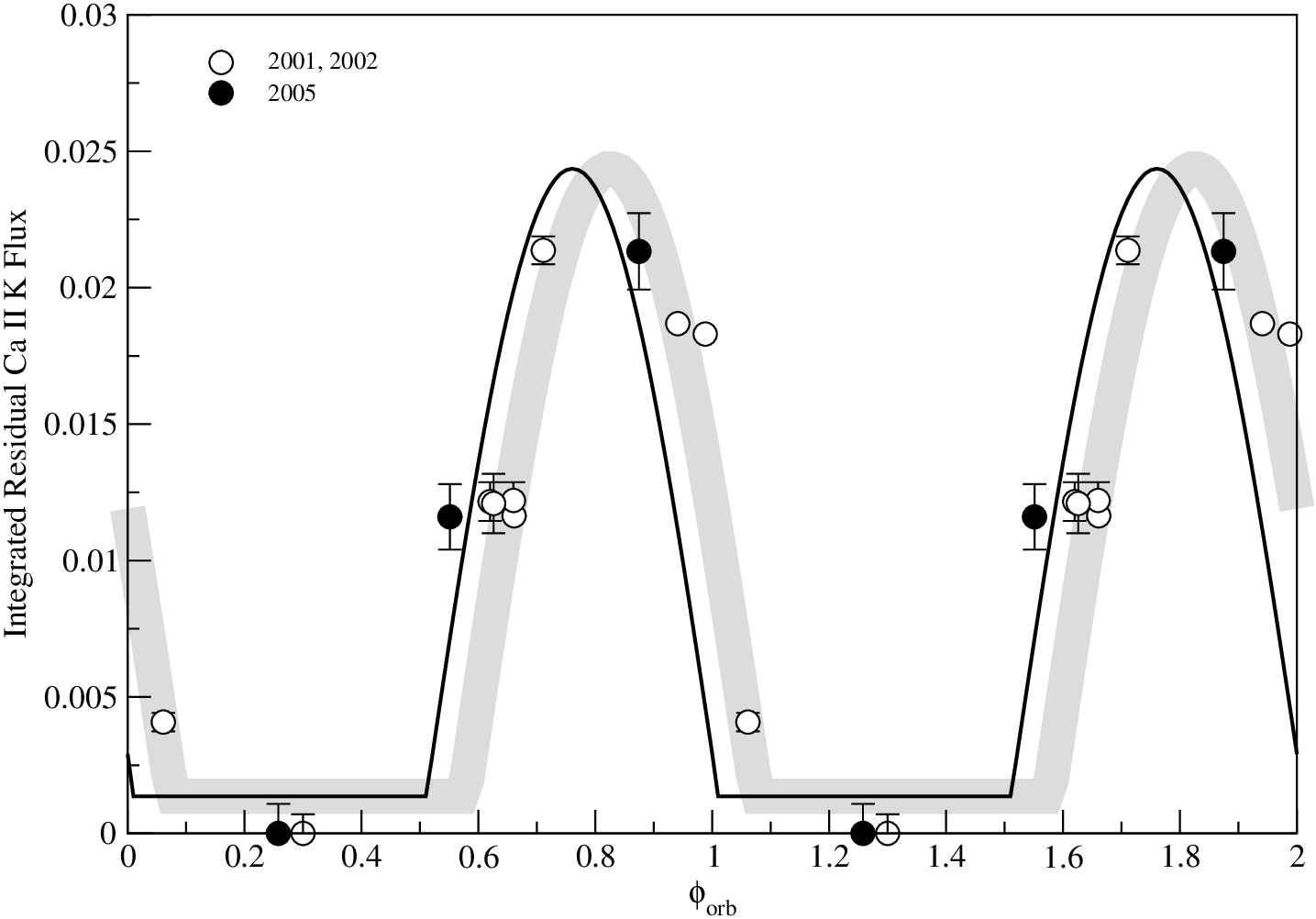}
\caption{Integrated flux of the K-line residuals from a normalized mean spectrum of HD~179949 as a function of orbital phase for the 2001 and 2002 data (open circles) published in Shkolnik et al.~(2003) and 2005 data (filled circles). The grey line is a best-fit spot model to the earlier data whose thickness reflects the error in the phase shift. The black line is the same fit slightly shifted in phase by -0.07 to better fit the 2005 data. This small shift relative to the earlier data is not signficant. Error bars in the integrated residual K flux are $\pm$ the intranight residual RMS. Note that the data points are repeated for two cycles.
\label{hd179949_intK}}
\end{figure}

In Shkolnik et al.~(2005a) we presented evidence of planet-induced heating on HD~179949. The effect lasted for over a year and
peaked only once per orbit, suggesting a magnetic interaction. In the simplest configuration, a magnetic interaction would occur
near the sub-planetary point, when the planet is in front of the star relative to the line-of-sight, which defines orbital phase
$\phi_{orb} = 0$. Reproduced in Figure~\ref{hd179949_intK}, we fitted a truncated, best-fit spot model to our 2001 and 2002 data
with $P = P_{orb}$ = 3.092 d corresponding to the change in projected area of a bright spot on the stellar surface before being
occulted by the stellar limb. The fit to the 2001 and 2002 data peaks at $\phi_{orb}$ = 0.83 $\pm$0.04 with an amplitude of
0.027. We over-plot new data from 2005 which is fit remarkably well by the same model with only an insignificantly small
relative phase shift of -0.07.

This phase lead may help identify the nature of the interaction.  For example, the offset from the sub-planetary point of a starspot or group of starspots can be a characteristic effect of tidal friction, magnetic drag or reconnection with off-center stellar magnetic field lines. For further discussion on such mechanisms, see papers by Gu et al.~(2005), Preusse et al.~(2006) and McIvor et al.~(2006). In any case, the phasing, amplitude and period of the activity have persisted for over 4 years.

\begin{figure}
\epsscale{.80}
\plotone{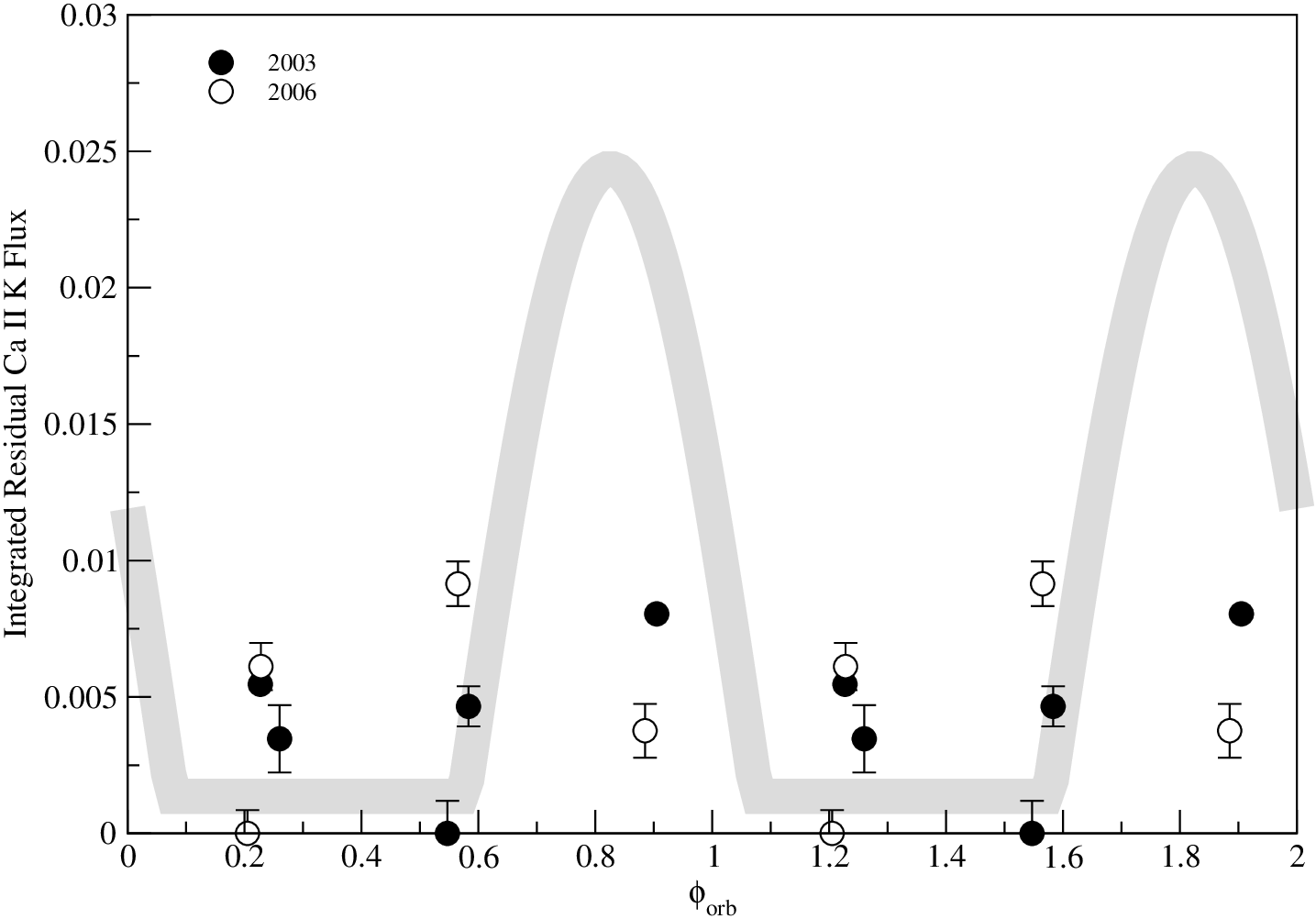}
\caption{Integrated flux of the K-line residuals from a normalized mean spectrum of HD~179949 for 2003 and 2006 data plotted on the 3.092-day orbital period with the SPI spot model from Figure~\ref{hd179949_intK} over-plotted.  Error bars are $\pm$ the intranight residual RMS. 
\label{hd179949_intK_2003_2006_orb}}
\end{figure}

\begin{figure}
\epsscale{.80}
\plotone{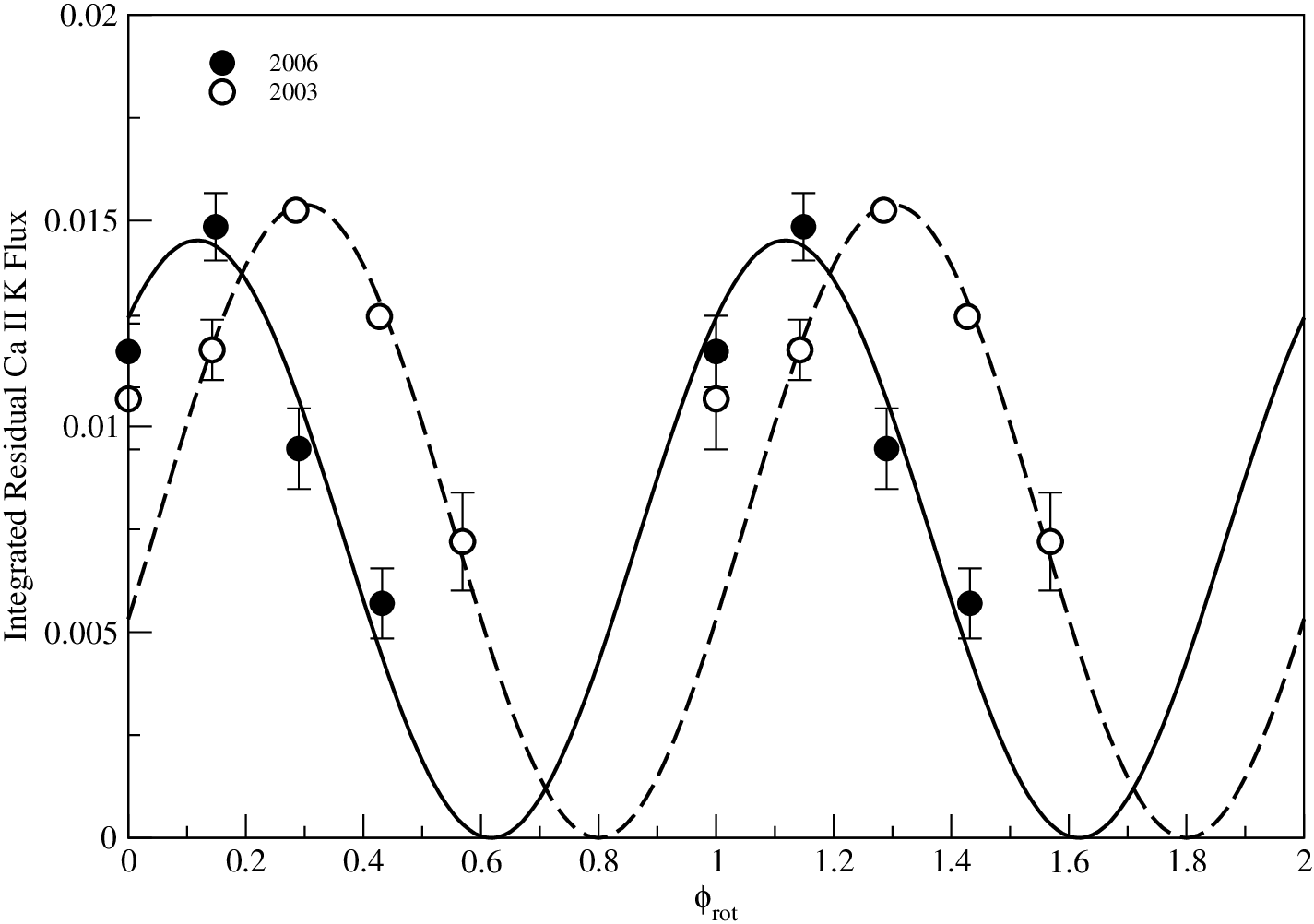}
\caption{Integrated flux of the K-line residuals from a normalized mean spectrum of HD~179949 for 2003 and 2006 data plotted on a 7-day rotation period with phases relative to the first night of each run. The points are vertically shifted such that the minimum of each curves is zero. Error bars are $\pm$ the intranight residual RMS. The curves are best-fit spot models to the two data sets. 
\label{hd179949_intK_rot}}
\end{figure}

Ca II data acquired in 2003 and 2006 of HD 179949 do not phase with the planet's orbit (Figure~\ref{hd179949_intK_2003_2006_orb}), but both phase well with a 7-day period, likely the rotation period of the star. In Figure~\ref{hd179949_intK_rot} we fit data from each year separately with a rotation curve because the effects of differential rotation and the appearance and disappearance of new spots over the three years would produce variations in phase, amplitude and period in the observed modulation. Note that the amplitude of the rotational activity in only 0.6 of that of induced by SPI.
Indirect indications of the rotation rate of HD 179949 imply $P_{rot} \approx 9$ days
and are presented in Shkolnik et al.~(2003) and Saar et al.~(2004). Wolf \& Harmanec (2004) weakly detect (1.5 $\sigma$) a photometric rotation period for HD~179949 of 7.07 d with an amplitude of only 0.008 mag.
While more photometry is needed to determine a rotation period conclusively, the modulated Ca II emission of this star in both 2003 and 2006 strongly suggests a rotation period of 7 days.

\begin{figure}
\epsscale{.80}
\plotone{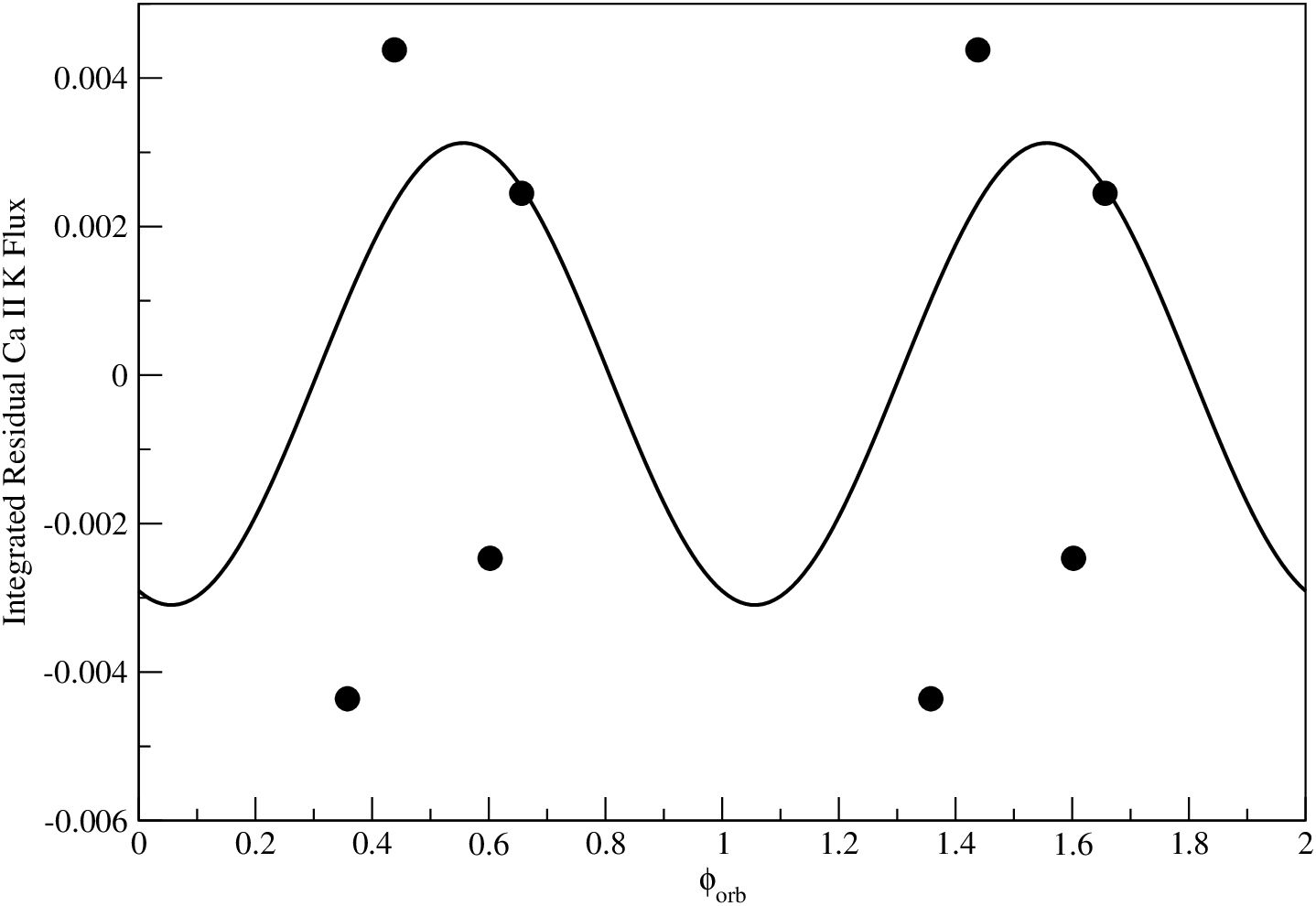}
\caption{Integrated flux of the K-line residuals from a normalized mean spectrum of $\upsilon$ And for September 2005 data plotted on the 4.6-day orbital period. Error bars are within the size of the of the points. The curve is the best-fit spot model to our 2002 and 2003 data. (See Figure 8 of Shkolnik et al.~2005.)
\label{upsAnd_intK_orb}}
\end{figure}

\begin{figure}
\epsscale{.80}
\plotone{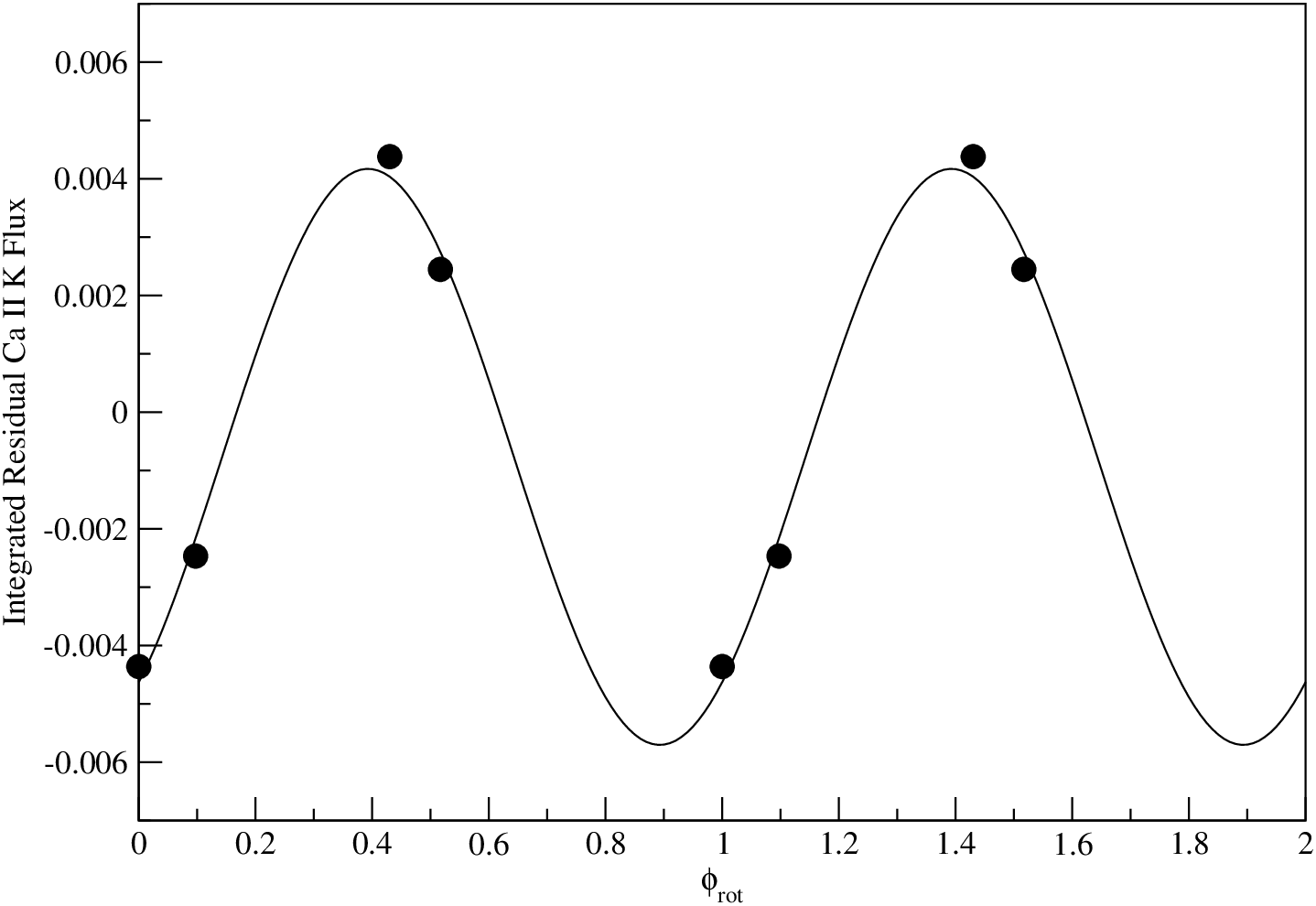}
\caption{Integrated flux of the K-line residuals from a normalized mean spectrum of $\upsilon$ And for September 2005 data plotted on a  12-day rotation period. Error bars are within the size of the of the points. The curve is a best-fit spot model.
\label{upsAnd_intK_rot}}
\end{figure}

Similarly, previous Ca II data of $\upsilon$ And indicated possible SPI (Figure 8 of Shkolnik et al.~2005a), yet our September 2005 data appears to vary with the rotation. Again, the rotation period is not well known. Henry et al.~(2000) quotes both 11 and 19 days, with a probable 11.6-day period from the $\langle S_{HK} \rangle$ index. We plot the 2005 data against the 4.6-d orbital in Figure~\ref{upsAnd_intK_orb} and an 11.6-day rotation period in Figure~\ref{upsAnd_intK_rot}. Unlike data from 2002 and 2003, the 2005 data phase much better with $P_{rot}$=11.6 d than with a planetary orbit.	

This on/off characteristic of SPI observed in the HD~179949 and $\upsilon$~And systems is predicted by the models of Cranmer \& Saar (2007). They model the Ca II H \& K light curve of a sun-like star with a hot Jupiter interacting with the field geometry at various stages of the empirically derived solar magnetic field at annual steps of the 11-year solar cycle. They conclude that due to the complex nature of the multipole fields, 
the Ca II K light curves due to SPI do not repeat exactly from orbit to orbit, and at times the planet-induced enhancement may disappear altogether leaving only rotationally modulated emission. This may explain the 2003 and 2006 disappearance of the strong orbital modulation seen in 2001, 2002 and 2005 for HD 179949. Their models also show that for sparsely sampled data, the apparent phase shift between the peak Ca II emission and the sub-planetary point may fall between −0.2 and +0.2 (or $\pm$ 72$^{\circ}$), consistent with the -0.17 phase shift we detected repeatedly for HD 179949.

\subsection{HD 189733: The active host of a massive planet}\label{hd189733}

\begin{figure}
\epsscale{.80}
\plotone{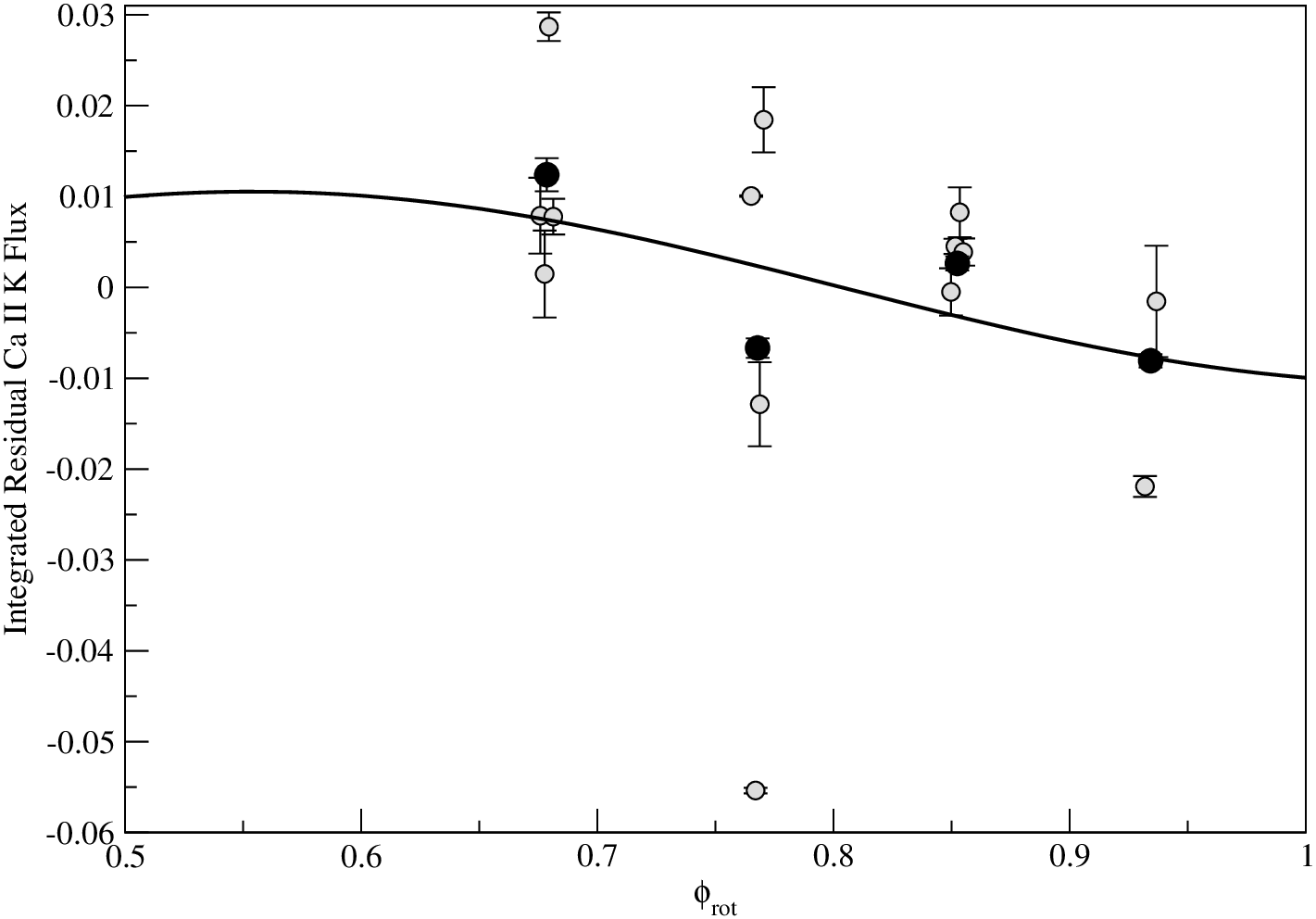}
\caption{Integrated flux of the K-line residuals from a normalized mean spectrum of HD~189733 as a function of a 11.7-day rotational phase (Moutou et al.~2007, Croll et al.~2007). The open circles are residuals of individual exposures relative to a global mean, while the solid circles use only the mean of each night. Error bars represent twice the measurement error of a given exposure as measured by the residuals outside of the Ca II K core. The solid curve is a best-fit sinusoid tracing the rotation of the star.  
\label{hd189733_intK_rot}}
\end{figure}

\begin{figure}
\epsscale{.80}
\plotone{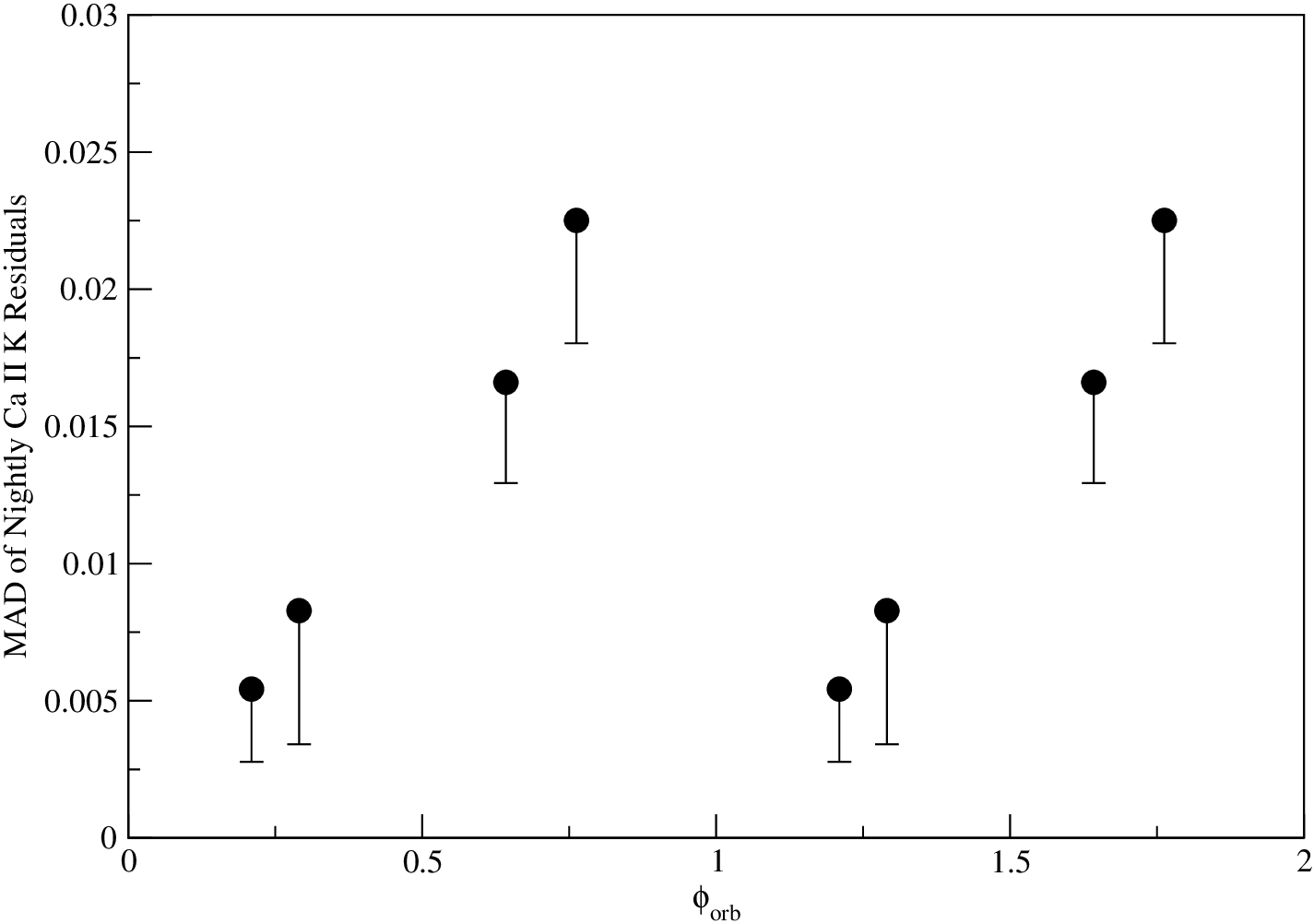}
\caption{Mean absolute deviation (MAD) of the nightly K-line residuals shown in Figure~\ref{hd189733_intK_rot} of HD~189733 as a function of the 2.2-day orbital period. The error bars represent the integrated MAD immediately outside the Ca II emission core and reflect the S/N obtained for each  night. Errors in phase are less than the size of the points.
\label{hd189733_MADK_orb}}
\end{figure}

We reported in Shkolnik et al.~2005a that chromospheric variability of the active, young star $\kappa^{1}$~Ceti ($\langle$K$\arcmin$$\rangle$=0.815), for which the presence of a hot Jupiter is not ruled out, and HD~73256, known to host a hot Jupiter (M$_p$sin$i$=1.85 M$_J$, $a$=0.037 AU, Udry et al.~2003, $\langle$K$\arcmin$$\rangle$=0.899) was modulated by stellar rotation with additional variability or flaring potentially induced by a hot Jupiter. We find a similar effect on HD 189733, a generally more active star, for which the average $P_{rot}$ is well known (11.73$\pm$0.07 days, Croll et al.~2007). This star has relatively strong K emission with $\langle$K$\arcmin$$\rangle$=1.231 and large intranight variability as shown in Figure~\ref{hd189733_intK_rot} which varies on time-scales at least as short as the length of the individual exposures (30 minutes). The average emission from night to night clearly varies with the star's rotation though with a lower amplitude (0.011 \AA) as compared to  HD~73256 (0.045 \AA), likely because the star is intrinsically more active, with a larger percentage of its surface covered in spots. This makes it difficult to extract any magnetic and/or tidal contribution to HD~189733's chromospheric emission by the planet as no correlation is seen between the planet's orbit and the residuals to the rotational modulation. However, Figure~\ref{hd189733_MADK_orb} shows how the integrated mean absolute deviation of the K line residuals from the global mean per night (MADK) vary with orbital phase. Though we only have four nights of observations spanning 1.4 orbits, there is a clear increase in very-short-term ($\leq$30 minutes) activity at $\phi_{orb}\sim0.8$. Remarkably, this is the same phase at which SPI peaks for HD 179949 and $\tau$~Boo as measured both spectroscopically and photometrically. (See discussion below and Walker et al.~2006, 2007)

\subsection{$\tau$ Boo: SPI on a tidally-locked star}\label{tauboo}

The star with the shortest rotation period in our sample is $\tau$~Boo. It has the largest $v$sin$i$ (= 14.8 m~s$^{-1}$, Gray 1982) and is
believed to be in synchronous rotation with its tightly-orbiting massive planet ($P_{rot}$ = 3.2 $\pm$ 0.5 d, Henry et al.~2000, $P_{orb}$ =
3.31250 d, M$_p$sin$i$=4.4 M$_J$, Butler et al.~1997). We observed a small but significant night-to-night modulation in the H \&
K emission of $\tau$ Boo during the first 3 years of observations with no obvious phasing with the planet's position. (Data from 2001, 2002 and
2003 are in Figure 7 of Shkolnik et al.~2005a.) Due to the tidal-locking of the star with the planet, we must depend on consistent orbital phasing of any modulated activity to disentangle SPI from rotational modulation for $\tau$ Boo.

\begin{figure}
\epsscale{.80}
\plotone{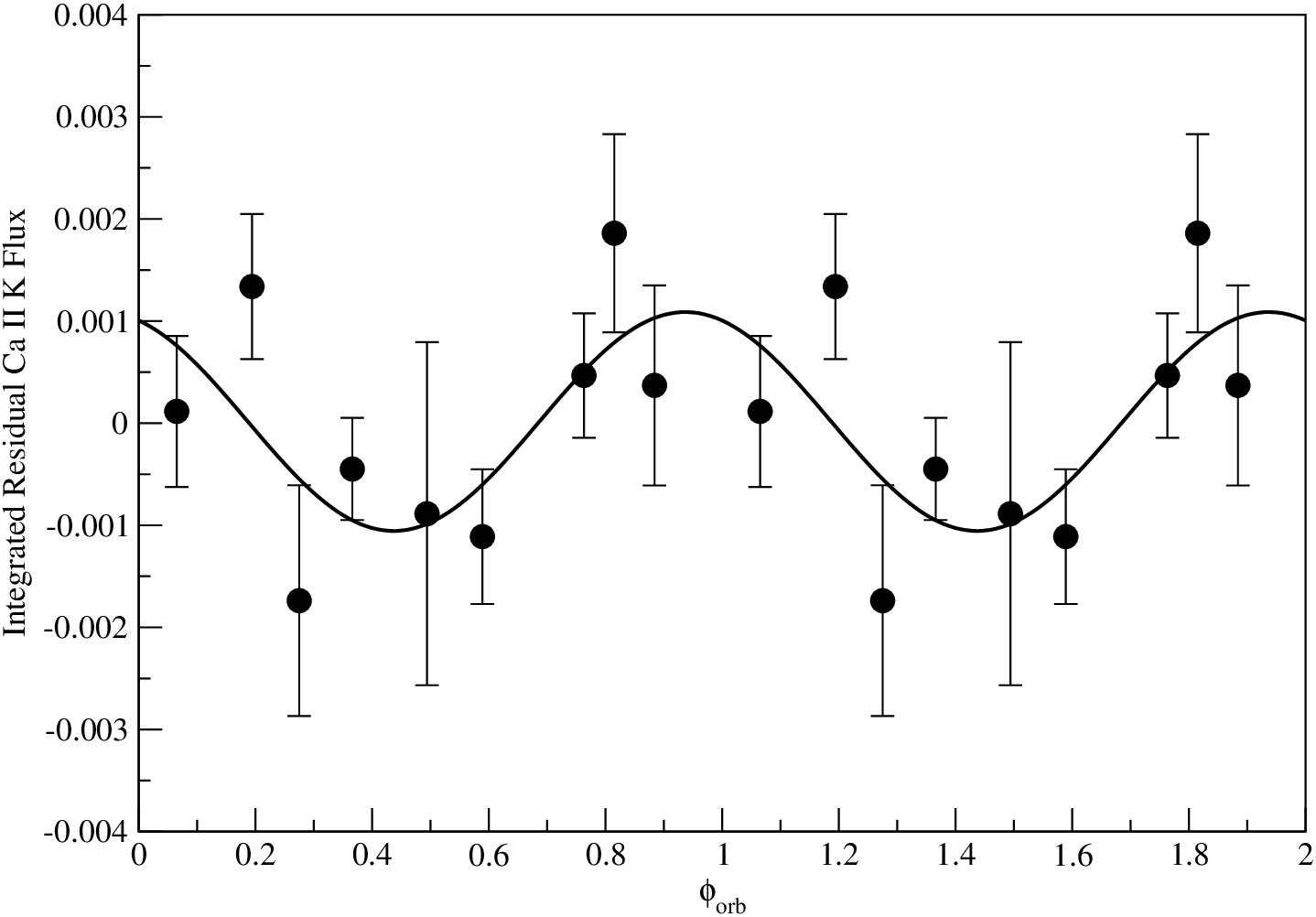}
\caption{Integrated flux of the K-line residuals from a normalized mean spectrum of $\tau$ Boo as a function of orbital phase with a best-fit spot model. Note that since $P_{rot} \simeq P_{orb}$ for this system, $\phi_{rot} \simeq \phi_{orb}$. The three observations taken in `spectropolarimetry'  mode are at $\phi_{rot}$ = 0.19, 0.49 and 0.82.  Error bars are $\pm$ the intranight residual RMS. 
\label{tauboo_intK}}
\end{figure}

Walker et al.~(2007) presented light curves of $\tau$ Boo taken in 2004 and 2005 observed in broadband optical light by the MOST space telescope. In the first year, they observed a significant photometric signal close to the planet's orbital frequency. In the second year, there was no signal of similar strength but a clear correlation with the MAD of the photometry with orbital phase. They showed that when phased to the planet's orbital period, the active region precedes the sub-planetary point by 68$^{\circ}$, very close to the phase lead we observe in the enhanced activity on HD~179949 and in the MAD of HD 189733's Ca II K emission. Though synchronisity with the exact planet's position is not obvious in $\tau$ Boo's 2001$-$2003 Ca II data, Walker et al.~(see their Figure 6) show that the MAD of these data during the photometrically active phase range, centered on $\phi_{orb}=$0.8, is twice as high than outside of it.  
Though of smaller amplitude (and larger error bars) than in previous years, our new 2006 Ca II data (Figure~\ref{tauboo_intK})\footnote{Of the 9 June 2006 nights on which $\tau$ Boo was observed, three were taken in `spectropolarimetry' mode, those at orbital phases of 0.19, 0.49 and 0.82. It is interesting to note that on the two nights with high K emission, Catala et al.~(2007) detect a clear Stokes V signature while the third night, ($\phi_{orb}$=0.49) has both low K emission and no Stoke V signal.} may also show a weak enhancement between $\phi_{orb}=$0.7 and 1.2. If this is indeed the case, it implies that an active region leading the sub-planetary point, has persisted on $\tau$ Boo for at least 5 years, equivalent to $\approx$550 planetary orbits.

\subsection{Night-to-night activity correlates with planet's magnetic moment}

\begin{figure}
\epsscale{0.65}
\plotone{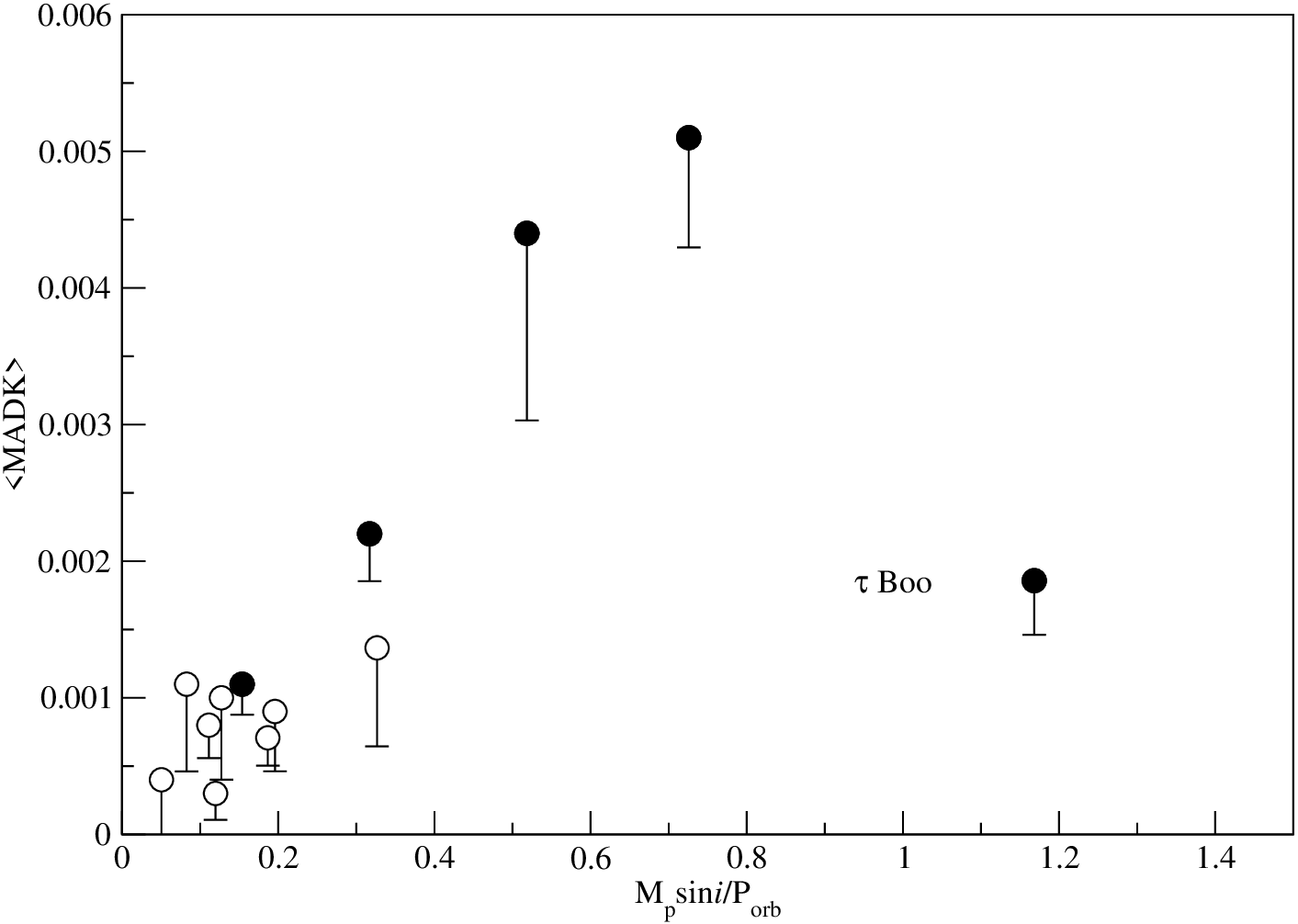}
\caption{The ratio of the minimum planetary mass (in Jupiter masses) to the orbital period (in days) plotted against the average MAD of the K-line per observing run for all 13 stars observed. The x-axis quantifies the planet's magnetic moment assuming tidal locking, such that $P_{rot}=P_{orb}$. The filled-in circles are of stars which exhibit significant night-to-night variability in the Ca II K line: HD 73526, $\upsilon$ And, HD 179949 and HD 189733 (this work). The tidally-locked system of $\tau$ Boo does not follow the correlation traced out by the others. The error bars are one-sided due to the positive contribution of integrated MAD immediately outside the Ca II emission core and reflect the S/N obtained for each target. 
\label{msini_MADK}}
\end{figure}

S\'anchez-Lavega (2004) looked at the internal structure and the convective motions of giant extrasolar planets in order to calculate their
dynamo-generated surface magnetism. Given the same angular frequency (which is a reasonable approximation for the short-period planets in
question), the magnetic dipole moment, and hence the magnetospheric strength, increases with planetary mass. This is observed for the magnetized planets in our own
solar system, where the magnetic moment grows proportionally with the mass of the planet (Stevens 2005), and more
specifically, with the planet's angular momentum ($L \propto$ $M_pR_p^2$$P_{p,rot}^{-1}$, Arge et al.~1995). Since only lower limits exist
for the masses of most hot Jupiters and at such small semi-major axes they should be tidally locked ($P_{p,rot}$ = $P_{orb}$), we plot
$M_{p}$sin$i$/P$_{orb}$ against $\langle$MADK$\rangle$, the average of the integrated MAD of the K line residuals per observing run, in
Figure~\ref{msini_MADK}. Though we are able to include only one additional $active$ point (HD 189733) to the original plot of Shkolnik et
al.~2005a, we continue to see an intriguing correlation between the planet's magnetic moment and the night-to-night chromospheric activity on
its star. Of our sample, $\tau$ Boo has the most massive planet and yet falls well below the correlation. This is consistent with the
proposed Alfv\'en wave model where the near zero relative motion due to the tidal locking of both the star and the planet ($P_{*,rot}$ =
$P_{p,rot}$ = $P_{orb}$) produces minimal SPI because of the weak Alfv\'en waves generated as the planet passes through the stellar
magnetosphere, thereby transporting little excess energy to the stellar surface along the magnetic field lines (Gu et al.~2005).
If this correlation between short-term activity and planetary magnetic moment holds for more hot Jupiter systems engaging in SPI, this could provide an empirical tool with which to estimate the strength of extrasolar planetary magnetic fields.

\subsection{Long-term stellar activity cycles}\label{long}

\begin{figure}
\epsscale{0.5}
\plotone{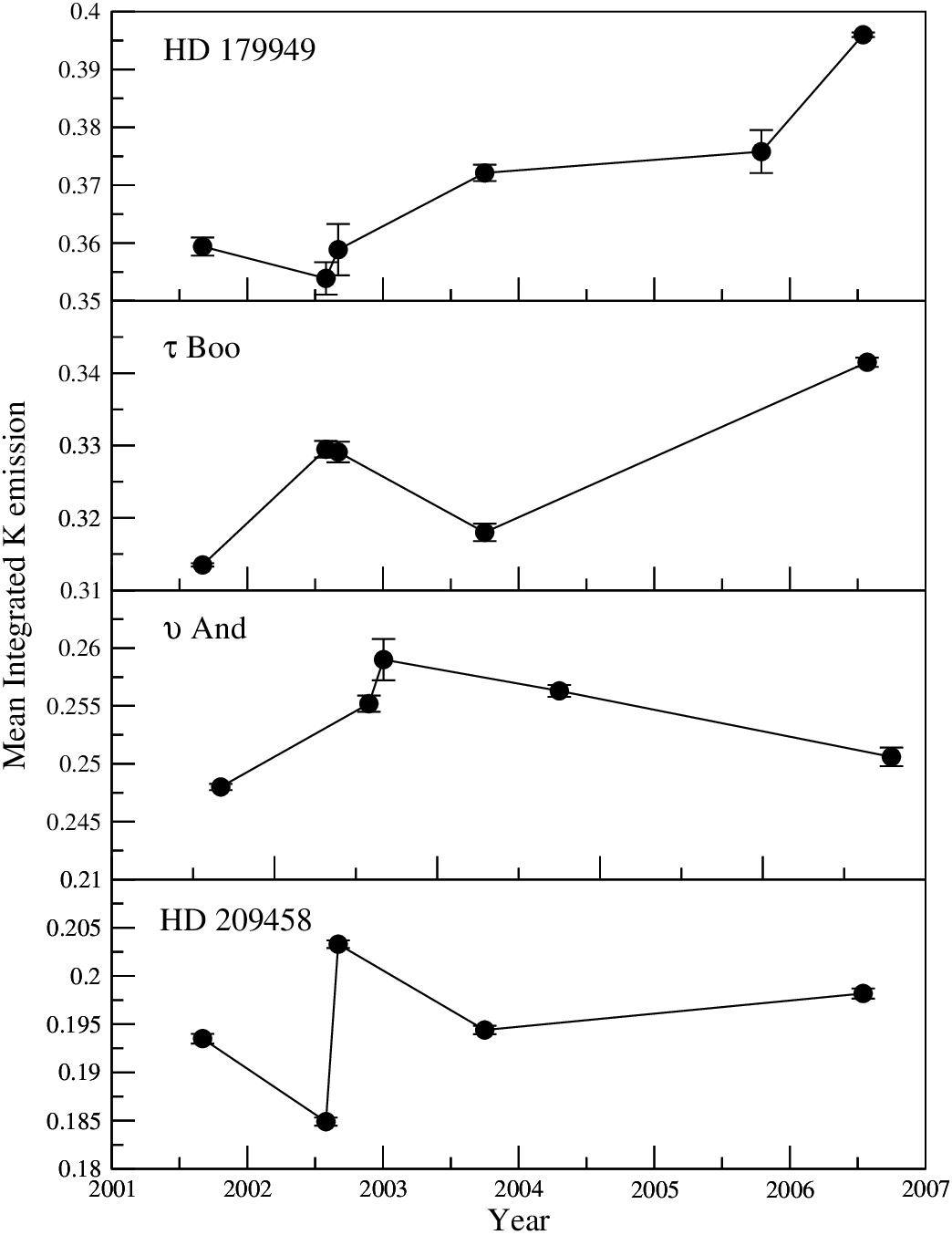}
\caption{The average integrated Ca II K emission of HD~179949, $\tau$ Boo, $\upsilon$ And and HD 209458 for each of the five or six observing runs. The error bars are the MAD values for each observing season.
\label{longterm_Kemission}}
\end{figure}

With CFHT data spanning 5 years, we can compare the long-term variations in the chromospheric level of HD 179949, $\tau$ Boo, $\upsilon$ And and HD~209458. We measure Ca II K emission strength $\langle$K$\rangle$ by integrating across the normalized K cores bounded by the K1 features (Montes et al.~1994) and plot their average for each observing run in Figure~\ref{longterm_Kemission}. The error bars represent the MAD value for each observing season. There does not appear to be any correlation between the mean chromospheric activity and the level of night-to-night modulation, be it due to SPI or stellar rotation.
Though 5-year baseline is a good start to tracking the intrinsic stellar activity cycles of these stars, limiting their periods to $>$10 years, the variability from run to run may also be due to active regions on the visible disk of the star. We require more frequent monitoring over several more years to firmly say anything more about the activity cycle of any individual program star.

\section{Correlating with other activity indicators}\label{other}

\begin{figure}
\epsscale{0.8}
\plotone{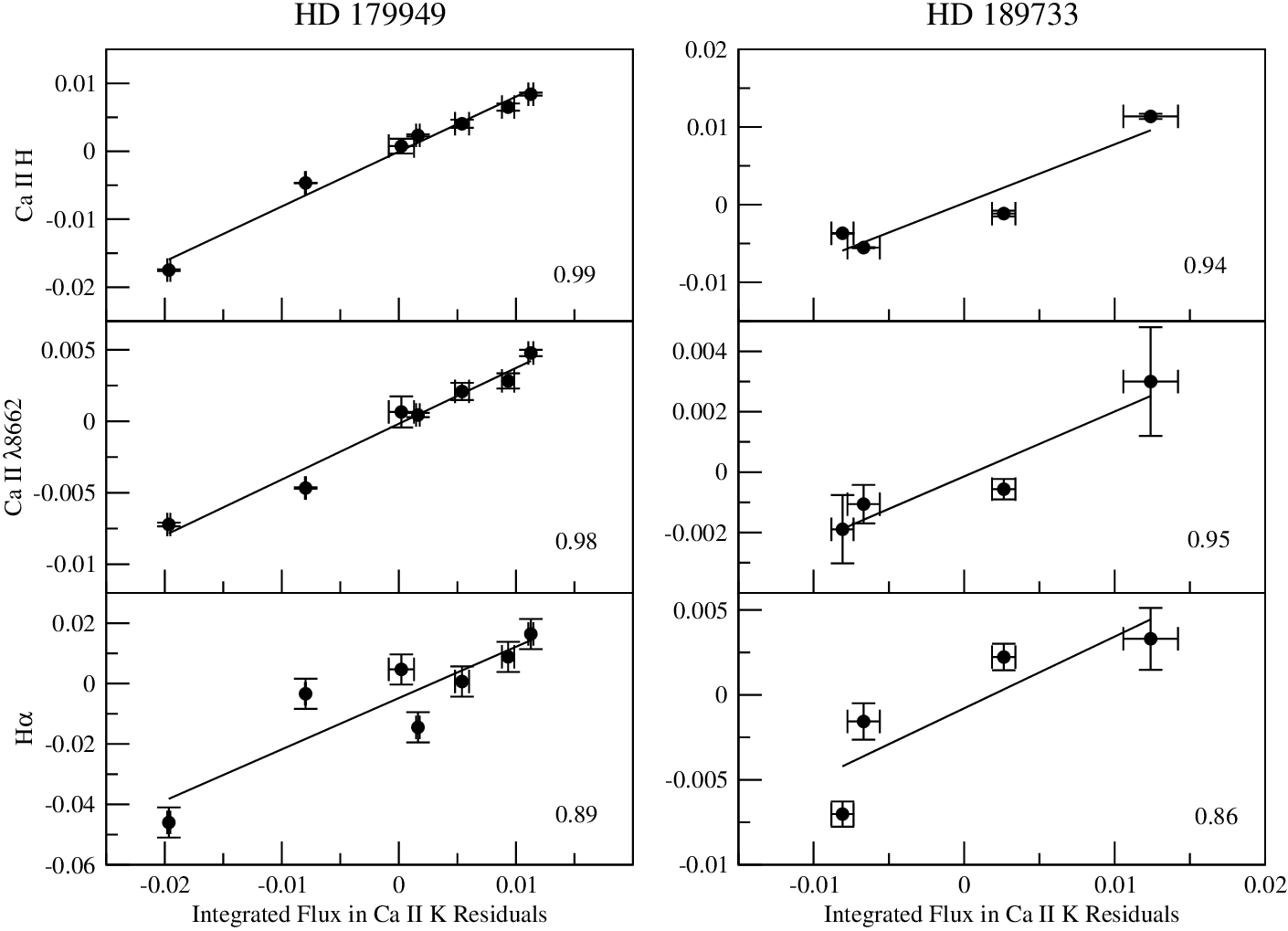}
\caption{Integrated flux in the Ca II K residuals plotted against the residual flux in Ca II H, the 8662\AA\/ line and H$\alpha$ for HD 179949 and HD 189733. The number in the corner is the correlation coefficient of the best-fit line.
\label{correlations}}
\end{figure}

A similar analysis to the Ca II K line described above was performed for the Ca II H, IRT line at 8662\AA\/ and H$\alpha$ though with the normalization points set at 0.7 of the local continuum for the latter two. This level is within the photospheric damping wings of the line, focusing the analysis on the chromospheric core and excluding some blended and telluric lines. For both HD 179949 and HD 189733, there is a strong correlation between the residuals of Ca II K with those of the Ca II H and 8662\AA\/ lines (Figure~\ref{correlations}) as expected given the common upper energy level of their transitions.  However, a poorer correlation exists with the H$\alpha$ lines. Though H$\alpha$ is often demonstrated to be just as good a tracer of chromospheric activity as Ca II, a recent analysis by Cincunegui et al.~(2007) has shown that when comparing a sample of stars, the correlation between H$\alpha$ and Ca II is the result of the correlation of each line with spectral type rather than with stellar activity. When comparing the variability of the lines for an individual star, there is no consistent correlation. This is likely the effect of the differing underlying formation physics between them. (Soderblom et al.~1993, Cram \& Mullan 1985). We leave the relative energy emitted in the lines and their implications for SPI models for a later paper.

The He I D3 line (a blended triplet) at 5876 \AA\/ correlates well with plage regions on the solar surface (Landman 1981) as well as  with the Ca II H\&K emission (Saar et al.~1997). It forms in the upper chromosphere and is thought to be back-heated by the stellar corona giving us a unique optical view of this hot plasma. The absence of He I absorption in non-magnetic regions on the sun and in other inactive stars indicates that He I has no basal (acoustically heated) flux level, unlike the other activity indicators in the visible spectrum, and is therefore purely a signature of magnetic activity.

\begin{figure}
\epsscale{0.8}
\plotone{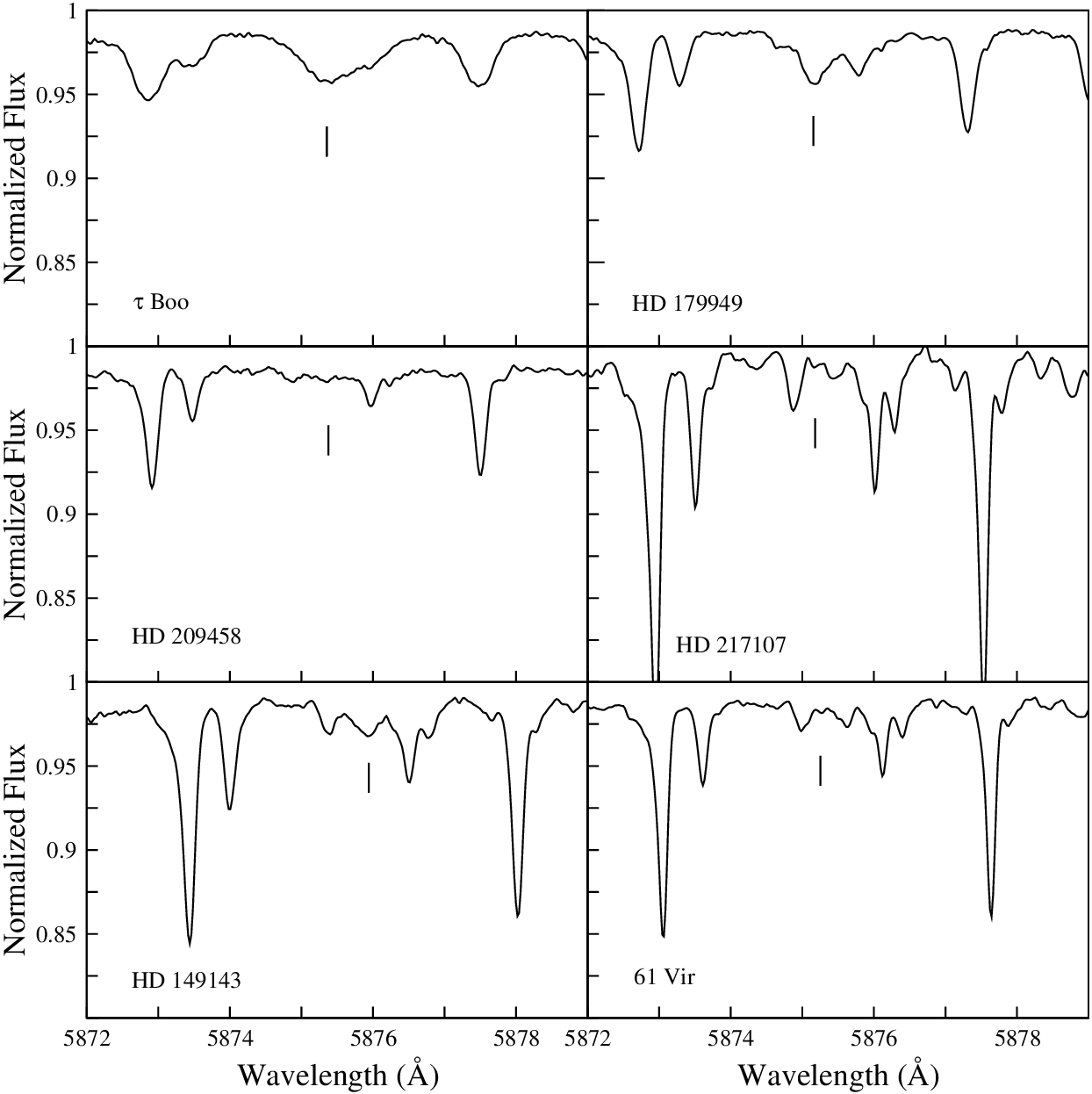}
\caption{Normalized spectra in the region of the He I line (5876 \AA) for 5 program stars plus the standard, 61 Vir. The vertical line corresponds to the location of the line. The same spectral region for HD 189733 is shown in Figure~\ref{hd189733_spec}. The He I cannot be measured for $\upsilon$~And since no telluric standard was observed on those nights.
\label{HeI_6stars}}
\end{figure}

We show the spectral region near the He I lines of our program stars in Figure~\ref{HeI_6stars}. The line is blended with the weak lines of Fe I 5876.30, Cr I 5876.55, and unidentified lines at 5875.76 and 5875.14 \AA, both blended with telluric H$_2$O (Moore et al.~1966).  The imperfect removal of the telluric lines from our spectra left residuals at the level of $\lesssim$0.003 of the nearby continuum. This made it difficult to analyze night-to-night variations in the stars that exhibit relatively strong variability in the Ca II lines, though the mere presence of the line indicates magnetic heating source, with strong absorption implying great activity.  HD~179949, HD~189733 and $\tau$~Boo all have clear He I absorption while HD~209458, HD~149143, HD~217107 as well as our standard star 61~Vir have only weak, if any, absorption.  Though night-to-night variability is difficult to quantify in HD 179949 and HD 189733, there is a clear increase in He I D3 absorption on the night that each of the two stars displays its maximum Ca II emission, which do not occur on the same night.\footnote{We cannot measure the He I EW of HD~179949 from the September 2005 data since poor weather prevented us from observing a telluric standard. Similarly, a telluric standard was not observed for the spectropolarimetric study of $\upsilon$ And.}

\begin{figure}
\epsscale{0.8}
\plotone{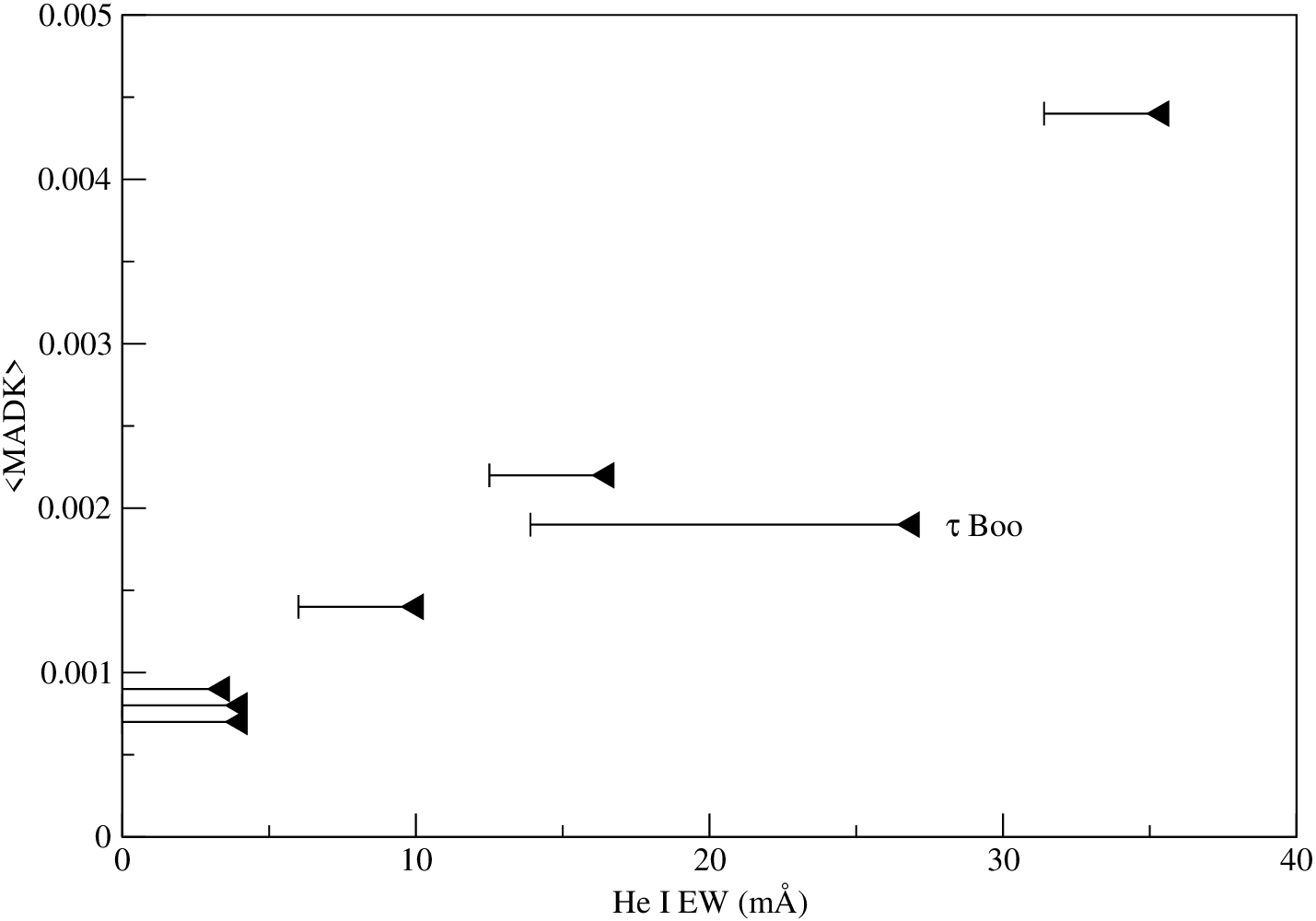}
\caption{Average MAD of the K emission $\langle$MADK$\rangle$ for the 2006 program stars as a function of the He I equivalent width. The error bars reflect the difficulty in measuring the EW due to line blends, especially for the rapidly rotating $\tau$~Boo. 
\label{madk_HeI}}
\end{figure}

We measured the average He I EW by deblending it with the contaminated lines, a technique particularly difficult for the rapid rotator $\tau$ Boo.  The values are listed in Table~\ref{targets}. It is interesting to note that the He I EW does not follow the power-law correlation with Ca II H \& K emission observed by Saar et al.~(1997) for G and K dwarfs, but does display a strong correlation with the short-term activity metric $\langle$MADK$\rangle$ plotted in Figure~\ref{madk_HeI}, where stars with more night-to-night activity have stronger He I absorption. The relationship between the He I absorption and stellar rotation period observed by Saar et al. is also not obvious in our small sample.  It therefore remains possible that the strength of the He I D3 line may predict whether a planet-bearing star will have night-to-night variability. When more stars with hot Jupiters are discovered this may come in handy in helping to decide which systems should be studied further with intensive time-series observations.

\section{Summary}\label{summary}

We have observed 7 stars with hot Jupiters using CFHT's \'echelle spectrograph ESPaDOnS to search for night-to-night modulation of the Ca II emission for evidence of SPI. Four of these have been observed in our previous studies of time-varying Ca II H \& K: HD 179949, $\upsilon$ And, HD 209458, and $\tau$~Boo, and we have added three new targets: HD 189733, HD 217107 and HD 149143. 

For our prime target, HD 179949, we now have a total of six observing runs spanning 5 years. During four runs (Aug 2001, July 2002, Aug
2002 and Sept 2005) the Ca II emission varied with the orbital period of 3.092 days, with consistent amplitude and peak phase
indicative of a magnetic interaction between the star and planet. The peak activity on HD 179949 in these epochs occurs at
$\phi_{orb}\approx0.8$, leading the sub-planetary longitude by some 70$^{\circ}$. Interestingly, this same phase shift is observed in the MAD of the Ca II K residuals of both $\tau$ Boo and HD~189733. The phase lead can provide information on the field
geometries (i.e.~Parker spiral) and the nature of the effect such as tidal friction, magnetic drag or reconnection with off-center magnetic fields.

HD 179949 data from the other two runs (Sept 2003 and June 2006) clearly vary with the rotation period of 7 days. A similar effect is seen on  $\upsilon$ And where one of four epochs appears to be modulated by rotation rather than the planet's motion. This {\it on/off}
behavior has been modeled by Cranmer \& Saar (2007) to be an effect of magnetic reconnection with the stellar field as it varies with
the star's long-term activity cycle. 

We present the expected correlations between the variability observed in the Ca II K line with Ca II H and IRT 8662, and a weaker correlation with H$\alpha$. Though we could not accurately measure the variability in the upper-chromosphere line He I D3, we show that it has the potential to flag stars which might be active in Ca II K in on a night-to-night time scale. 

To date we have observed 13 stars with hot Jupiters at CFHT and VLT, of which 5 appear to be actively engaging in SPI: HD 179949, $\upsilon$
And, HD 189733, HD 73256 and $\tau$ Boo. The activity as measured by the mean absolute deviation over a run on the first four of these stars
correlates well with $M_{p}$sin$i$/$P_{orb}$, a value proportional to the planet's magnetic moment, and thus with the hot Jupiter's magnetic
field strength. Because of their small separation ($\leq$ 0.075 AU), a hot Jupiter lies within the Alfv\'en radius of its host star,
allowing a direct magnetic interaction with the stellar surface. 
Although this correlation is tentative, short-term chromospheric variability
may be our first probe of extrasolar planetary magnetospheres.

\acknowledgements

We are very grateful to Claude Catala and John Barnes for contributing several $\tau$ Boo and $\upsilon$ And spectra, to Jean-Francois Donati for making {\it Libre Esprit} available to CFHT users, to the CFHT staff for their excellent support of this program, and to Benjamin Brown for useful discussions on planetary dynamos. Research funding from the NASA Postdoctoral Program (formerly the NRC Research Associateship) for E.S. and the National Research Council of Canada (D.A.B.) are gratefully acknowledged.

\clearpage


\begin{thebibliography}

\bibitem[]{413}Arge, C.N., Mullan, D.J., Dolginov, A.Z., 1995, ApJ, 443, 795

\bibitem[]{415}Bakos, G. \'A., Knutson, H., Pont, F., Moutou, C., Charbonneau, D., Shporer, A., Bouchy, F., Everett, M., Hergenrother, C., Latham, D. W., Mayor, M., Mazeh, T., Noyes, R. W., Queloz, D., P\'al, A., Udry, S., 2006, ApJ, 650, 1160

\bibitem[]{417}Bouchy F., Udry S., Mayor M., Moutou C., Pont F., Iribarne N., Da Silva R., Llovaisky S., Queloz D., Santos N.C., Segransan D., Zucker S., 2005, A\&A, 444, L15

\bibitem[Butler et al.(1997)]{But97} Butler, R.P., Marcy G.W., Williams, E., Hauser, H., Shirts, P., 1997, \apjl, 474, L115

\bibitem[]{421}Catala, C., Donati, J.-F., Shkolnik, E., Bohlender, D., Alecian, E., 2007, MNRAS, 374, 42

\bibitem[]{423}Catalano, S., Rodon\`o, M., Frasca, A., Cutispoto, C., 1996, IAUS, 176, 403

\bibitem[Charbonneau et al.(1999)]{Cha99}Charbonneau, D., Brown, T., Latham, D., Mayor, M., Mazeh, T., 1999, \apj, 529, 45

\bibitem[]{427}Cincunegui, C., D\`iaz, R.F., Mauas, P.J.D., 2007, A\&A, 469, 309


\bibitem[]{430}Cram, L.E., Mullan, D.J., 1985, ApJ, 294, 626

\bibitem[]{432}Cranmer, S. \& Saar, S., 2007, arXiv:astro-ph/0702530v1

\bibitem[]{434}Croll, B., Matthews, J., Walker, G., 2007, ApJ, submitted

\bibitem[]{436}Cumming, A., Marcy, G.W., Butler, R.P., 1999, ApJ, 526, 890

\bibitem[Cuntz et al. 2000]{Cun00} Cuntz, M., Saar, S.H., Musielak, Z.E., 2000, \apj, 533, L151

\bibitem[]{440}De Silva et al.~2006, A\&A, 446, 717

\bibitem[]{442}Donati, J.-F., Semel, M., Carter, B. D., Rees, D. E., Collier Cameron, A. 1997, MNRAS, 291, 658 

\bibitem[]{444}Donati, J.-F. 2007, MNRAS, in preparation

\bibitem[]{446}Fischer D., Marcy G., Butler P., Vogt S., Apps K., 1998, PASP, 111, 50

\bibitem[]{448}George, Samuel, Stevens, Ian, 2007, arXiv:0708.4079v1 

\bibitem[]{450}Glebocki, R., Bielicz, E., Pastuszka, Z., Sikorski, J., 1986, AcA, 36, 369

\bibitem[Gray 1982]{Gra82} Gray, D.F., 1982, \apj, 255, 200
	
\bibitem[]{454}Griessmeier, J.-M. Stadelmann, A., Penz, T., Lammer, H., Selsis, F., Ribas, I., Guinan, E. F., Motschmann, U., Biernat, H. K. Weiss, W. W., 2004, A\&A, 425, 753

\bibitem[]{456}Gu, P.-G., Shkolnik, E., Li, S.-L., Liu, X-W, 2005, AN, 326, 909


\bibitem[]{459}Henry, G.W., Baliunas, S.L., Donahue, R.A., Fekel, F.C., Soon, W., 2000, ApJ, 531, 415

\bibitem[]{461}Ip, W.-H., Kopp, A., Hu, J.-H., 2004, ApJ, 602, L53


\bibitem[]{464}Kashyap, V., Drake, J.,Steve Saar, S., 2006, Cool Stars 14 - Abstract \# 208

\bibitem[]{466}Khodachenko, M. L., Lammer, H., Lichtenegger, H. I. M., Langmayr, D., Erkaev, N. V., Grießmeier, J.-M., Leitner, M., Penz, T., Biernat, H. K., Motschmann, U., Rucker, H. O., 2007, P\&SS, 55, 631

\bibitem[]{468}Landman, D.A., 1981, ApJ, 251, 768

\bibitem[]{470}Lazio, T.J.W., Farrell, W.M., 2007, arXiv:0707.1827v1

\bibitem[Mazeh et al. 2000]{Maz00} Mazeh, T., Naef, D., Torres, G., Latham, D.W., Mayor, M., Beuzit, J.-L., Brown, T.M., Buchhave, L., Burnet, M., Carney, B.W., Charbonneau, D., Drukier, G., Laird, J.B., Pepe, F., Perrier, C., Queloz, D., Santos, N.C., Sivan, J.-P., Udry, S., Zucker, S., 2000, \apj, 532, L55

\bibitem[]{474}	
	McIvor, T., Jardine, M., Holzwarth, V., 2006, MNRAS, 367, 1


\bibitem[]{478}Montes, D., Fernandez-Figueroa, M.J., de Castro, E., Cornide, M., 1994, A\&A, 285, 609


\bibitem[]{481}Moutou, C., Donati, J.-F., Savalle, R., Hussain, G., Alecian, E., Bouchy, F., Catala, C., Collier Cameron, A., Udry, S., Vidal-Madjar, A., 2007, A\&A, 473, 651

\bibitem[Noyes et al. 1984]{Noy84} Noyes, R.W., Hartmann, L.W., Baliunas, S.L., Duncan, D.K., Baughan, A.H., 1984, \apj, 279, 763

\bibitem[]{485}	
	Olson, P., Christensen, U. R., 2006, E\&PSL. 250, 561


\bibitem[]{489}Pont, F., Bouchy, F., Melo, C., Santos, N. C., Mayor, M., Queloz, D., Udry, S., 2005, A\&A, 438, 1123
	
\bibitem[]{491}Preusse, S., Kopp, A., B\"uchner, J., Motschmann, U. 2006, A\&A, 460, 317 

\bibitem[]{493}Saar, S. H., Huovelin, J., Osten, R. A., Shcherbakov, A. G. 1997, A\&A 326, 741

\bibitem[]{495}Saar, S.H., Cuntz, M., Shkolnik, E., 2004, in IAU Symp. 219, Stars as Suns: Activity, Evolution, Planets, eds. A.K. Dupree and A.O. Benz, p.355

\bibitem[]{497}Saar, S., Kashyap, V., Cuntz, M., Shkolnik, E., Hall, J., 2006 Cool Stars 14 Meeting - Abstract \# 252

\bibitem[]{499}S\'anchez-Lavega, A., 2004, ApJ, 609, 87

\bibitem[Shkolnik et al.(2003)]{Shk03}
    Shkolnik, E., Walker, G.A.H.,
    Bohlender, D.A.,
    2003, \apj, 597, 1092

\bibitem[Shkolnik et al.(2005)]{Shk05_01}
    Shkolnik, E., Walker, G.A.H.,
    Bohlender, D.A.,Gu, P.-G., K\"urster, M., 2005a, ApJ, 622, 1075

\bibitem[Shkolnik et al.(2005)]{Shk05_02}
    Shkolnik, E., Walker, G.A.H., Rucinski, S.,
    Bohlender, D.A., Davidge, T.J., 2005b, AJ, 130, 799 

	
\bibitem[]{515}Soderblom, D.R., Stauffer, J.R., Hudon, J.D., Jones, B.F., 1993, ApJS, 85, 315

\bibitem[]{517}Stevens, I. R. 2005, MNRAS, 356, 1053

\bibitem[Tinney et al.(2001)]{Tin01}Tinney, C., Butler, P., Marcy, G., Jones, H., Penny, A., Vogt, S., Apps, K., Henry, C., 2001, \apjl, 551, L507

\bibitem[]{521}Udry, S., Mayor, M., Clausen, J.V., Freyhammer, L.M., Helt, B.E., Lovis, C., Naef, D., Olsen, E.H., Pepe, F., Queloz, D., Santos, N.C., 2003, A\&A, 407, 679

\bibitem[]{523}	
	Valenti, J. A., Fischer, D. A., 2005, ApJS, 159, 141

\bibitem[Vidal-Madjar et al 2003]{Vid03} Vidal-Madjar, A., Lecavelier des
\'{E}tangs, A., D\'{e}sert, J.-M., Ballester, G., Ferlet, R., H\'{e}brard, G., Mayor, M., 2003, Nature, 422, 143

\bibitem[]{529}	
	Vidal-Madjar, A., D\'esert, J.-M., Lecavelier des Etangs, A., H\'ebrard, G., Ballester, G. E., Ehrenreich, D., Ferlet, R., McConnell, J. C., Mayor, M., Parkinson, C. D., 2004, ApJ, 604, L69

\bibitem[Walker et al.(2003b)]{MOST}
    Walker, G.A.H., Matthews, J.M., Kuschnig, R., Johnson, R.,
    Rucinski, S., Pazder, J., Burley, G., Walker, A., Skaret, K.,
    Zee, R., Grocott, S., Carroll, K., Sinclair, P., Sturgeon, D.,
    Harron, J.,
    2003b, \pasp, 115, 1023

\bibitem[]{539}Walker G. A. H., Matthews, J.M., Kuschnig, R., Rowe, J.F., Guenther, D. B., Moffat, A. F. 
	J., Rucinski, S.M., Sasselov, D., Seager, S.,Shkolnik, E., Weiss, W. W. 2006, in Arnold 
	L., Bouchy F., Moutou C., eds, Proc. Haute-Provence Observatory Coll., Tenth An- 
	niversary of 51 Peg- B: Status of and Prospect for Hot Jupiter Studies. Frontier Group, 
	Paris, p. 267 (http://www.obs-hp.fr/www/pubs/Coll51Peg/proceedings.html) 

\bibitem[]{545}Walker, G.A.H, Croll, B., Matthews, J.M., Kuschnig, R., 
	Huber, D., Weiss, W.W., Shkolnik, Rucinski, S.M., Guenther, D.B., 
	Moffat, A.F.J., Sasselov, D., 2007, ApJ, submitted

\bibitem[]{549}Wolf, M., Harmanec, P., 2004, IBVS, 5575, 1


\bibitem[]{552}Wright, J., Marcy, G., Butler, R., Vogt, S., 2004, ApJS, 152, 261


\bibitem[]{555} Zarka, P., Treumann, R.A., Ryabov, B.P., Ryabov, V.B., 2001, Ap\&SS, 277, 293

\end{thebibliography}
\end{document}